\documentstyle[psfig]{mn}

\newif\ifAMStwofonts
\AMStwofontstrue

\ifoldfss
  \ifCUPmtlplainloaded \else
    \NewTextAlphabet{textbfit} {cmbxti10} {}
    \NewTextAlphabet{textbfss} {cmssbx10} {}
    \NewMathAlphabet{mathbfit} {cmbxti10} {} % for math mode
    \NewMathAlphabet{mathbfss} {cmssbx10} {} %  "   "    "
  \fi
  \ifAMStwofonts
    \ifCUPmtlplainloaded \else
      \NewSymbolFont{upmath} {eurm10}
      \NewSymbolFont{AMSa} {msam10}
      \NewMathSymbol{\upi}     {0}{upmath}{19}
      \NewMathSymbol{\umu}     {0}{upmath}{16}
      \NewMathSymbol{\upartial}{0}{upmath}{40}
      \NewMathSymbol{\leqslant}{3}{AMSa}{36}
      \NewMathSymbol{\geqslant}{3}{AMSa}{3E}

    \fi
  \fi
\fi % End of OFSS

\ifnfssone
  \newmathalphabet{\mathit}
  \addtoversion{normal}{\mathit}{cmr}{m}{it}
  \addtoversion{bold}{\mathit}{cmr}{bx}{it}
  \newmathalphabet{\mathbfit} % math mode version of \textbfit{..}
  \addtoversion{normal}{\mathbfit}{cmr}{bx}{it}
  \addtoversion{bold}{\mathbfit}{cmr}{bx}{it}
  \newmathalphabet{\mathbfss} % math mode version of \textbfss{..}
  \addtoversion{normal}{\mathbfss}{cmss}{bx}{n}
  \addtoversion{bold}{\mathbfss}{cmss}{bx}{n}
  \ifAMStwofonts
    \ifCUPmtlplainloaded \else
      \UseAMStwoboldmath
      \makeatletter
      \new@mathgroup\upmath@group
      \define@mathgroup\mv@normal\upmath@group{eur}{m}{n}
      \define@mathgroup\mv@bold\upmath@group{eur}{b}{n}
      \edef\UPM{\hexnumber\upmath@group}
      \new@mathgroup\amsa@group
      \define@mathgroup\mv@normal\amsa@group{msa}{m}{n}
      \define@mathgroup\mv@bold\amsa@group{msa}{m}{n}
      \edef\AMSa{\hexnumber\amsa@group}
      \makeatother
      \mathchardef\upi="0\UPM19
      \mathchardef\umu="0\UPM16
      \mathchardef\upartial="0\UPM40
      \mathchardef\leqslant="3\AMSa36
      \mathchardef\geqslant="3\AMSa3E
    \fi
  \fi
\fi % End of NFSS release 1

\ifnfsstwo
  \DeclareMathAlphabet{\mathbfit}{OT1}{cmr}{bx}{it}
  \SetMathAlphabet\mathbfit{bold}{OT1}{cmr}{bx}{it}
  \DeclareMathAlphabet{\mathbfss}{OT1}{cmss}{bx}{n}
  \SetMathAlphabet\mathbfss{bold}{OT1}{cmss}{bx}{n}
  \ifAMStwofonts
    \ifCUPmtlplainloaded \else
      \DeclareSymbolFont{UPM}{U}{eur}{m}{n}
      \SetSymbolFont{UPM}{bold}{U}{eur}{b}{n}
      \DeclareSymbolFont{AMSa}{U}{msa}{m}{n}
      \DeclareMathSymbol{\upi}{0}{UPM}{"19}
      \DeclareMathSymbol{\umu}{0}{UPM}{"16}
      \DeclareMathSymbol{\upartial}{0}{UPM}{"40}
      \DeclareMathSymbol{\leqslant}{3}{AMSa}{"36}
      \DeclareMathSymbol{\geqslant}{3}{AMSa}{"3E}
    \fi
  \fi
\fi % End of NFSS release 2

\ifCUPmtlplainloaded \else
  \ifAMStwofonts \else % If no AMS fonts
    \def\upi{\pi}
    \def\umu{\mu}
    \def\upartial{\partial}
  \fi
\fi

\title[Near and mid-infrared colours of star-forming galaxies]
{Near and mid-infrared colours of star-forming galaxies in European Large 
Area {\em ISO} Survey fields}
\author[P. V\"ais\"anen et al.]
       {P. V\"ais\"anen$^{1,2,}$\thanks{E-mail: pvaisane@eso.org}, 
        T. Morel$^{3,4,5}$, M. Rowan-Robinson$^{3}$, S. Serjeant$^{3}$,
        S. Oliver$^{3,6}$,
          \vspace*{0.2cm} \\
       {\LARGE T. Sumner$^{3}$, H. Crockett$^{3}$, 
        \vspace*{0.2cm} C. Gruppioni$^{7,8}$, E.\ V. Tollestrup$^{9,10}$} \\
        $^{1}$ Observatory, P.O.Box 14, University of Helsinki, Finland\\
        $^{2}$ European Southern Observatory, Alonso de Cordova 3107, Vitacura, Casilla 19001, Santiago 19, Chile\\
        $^{3}$ Astrophysics Group, Blackett Laboratory, Imperial College of
Science Technology \& Medicine, Prince Consort
Rd.,London.SW7 2BZ\\
        $^{4}$ IUCAA, Post Bag 4 Ganeshkhind Pune 411 007 India \\
        $^{5}$ Osservatorio Astronomico di Palermo, Piazza del Parlamento 1, 90134 Palermo, Italy\\
        $^{6}$ Astronomy Centre, Physics and Astronomy Subject Group, School 
of CPES, University of Sussex, Falmer, Brighton BN1 9QJ \\
        $^{7}$ Osservatorio Astronomico di Padova, vicolo dell'Osservatorio 5, 35122 Padova, Italy\\
        $^{8}$ Osservatorio Astronomico di Bologna, via Ranzani 1,
40127 Bologna, Italy\\
        $^{9}$ Boston University, Department of Astronomy, 725 Commonwealth 
Avenue, Boston, MA 02215, USA \\
        $^{10}$ Harvard-Smithsonian Center for Astrophysics, 60 Garden 
Street,
Cambridge, MA 02138, USA }

\pagerange{\pageref{firstpage}--\pageref{lastpage}}
\pubyear{2001}

\begin{document}

\maketitle

\label{firstpage}

\newcommand{\ea}{{et~al.~}}
\newcommand{\eg}{{e.g.~}}
\newcommand{\ie}{{i.e.~}}

\begin{abstract}

We present $J$ and $K$-band near-infrared photometry of a sample of
mid-infrared sources detected by the {\em Infrared Space Observatory}
({\em ISO}) as part of the European Large Area {\em ISO}-Survey (ELAIS)
and study their classification and star-forming properties.
We have used the Preliminary ELAIS Catalogue for the 6.7 $\umu$m (LW2) and 
15 $\umu$m (LW3) fluxes.
All of the high-reliability LW2 sources and 80 per cent of the LW3 sources
are identified in the near-IR survey reaching $K \approx 17.5$ mag.
The near- to mid-IR flux ratios can effectively be used to separate
stars from galaxies in mid-IR surveys.
The stars detected in our survey region are
used to derive a new accurate calibration for the ELAIS ISOCAM data in both 
the LW2 and LW3 filters.   We show that near to mid-IR
colour-colour diagrams can be used to further classify galaxies, as well as
study star-formation.  The ISOCAM ELAIS survey is
found to mostly detect strongly star-forming late-type galaxies, 
possibly starburst powered galaxies, and it also
picks out obscured AGN.
The ELAIS galaxies yield an average mid-IR flux ratio LW2/LW3 $=0.67 \pm 0.27$.
We discuss the $f_{\nu}(6.7\umu{\rm m}) / f_{\nu}(15\umu{\rm m})$ 
ratio as a star formation tracer using {\em ISO} and 
{\em IRAS} data of a local comparison sample.  We find that the 
$f_{\nu}(2.2\umu{\rm m}) / f_{\nu}(15\umu{\rm m})$ ratio is also a
good indicator of activity level in galaxies and conclude
that the drop in the $f_{\nu}(6.7\umu{\rm m}) / f_{\nu}(15\umu{\rm m})$ ratio
seen in strongly star-forming galaxies is a result of both an increase of 
$15\umu$m emission and an apparent depletion of $6.7\umu$m emission.
Near-IR data together with the mid-IR give the possibility to estimate 
the relative amount of interstellar matter in the galaxies.

\end{abstract}

\begin{keywords}
infrared: galaxies -- galaxies: evolution -- galaxies:
star-burst -- surveys -- infrared: stars
\end{keywords}

\section{Introduction}

There has been determined effort over the past several years to understand
the history of luminous matter in the Universe.
Ultimately, one wishes to have a consistent understanding which would tie
together the detailed physical processes at work in stars and ISM
in the Milky Way and local galaxies with the integrated properties of
more distant
systems.  The spectral properties and energy budget of
the distant galaxies in turn are
crucial in understanding the universal history of star formation,
the very faintest source counts, and the extragalactic background radiation.

In particular, the infrared and sub-mm regimes
have become the focal point of interest in studies of galaxies, both normal
and extreme objects.
The near-infrared is an important region for galaxy evolution studies for
several reasons.   Dust extinction is significantly
less hampering here than in the optical, and the light
mostly comes from a relatively stable old population of late-type stars
making galaxy colours, counts, and K-corrections easier to predict and
interpret.
It is also in the near-IR that the energy output of a galaxy
starts to shift from normal starlight to emission re-radiated by
interstellar matter. By $5 \umu$m the dust emission has taken
over from radiation from stellar photospheres, except in most ellipticals.

Apart from the $[12/25] \equiv f_{\nu}(12\umu{\rm m}) / f_{\nu}(25\umu{\rm 
m})$
colours of {\em IRAS} galaxies,
the mid-infrared truly opened up for study only with the {\em ISO}-mission
(see reviews by Genzel \& Cesarsky 2000, Helou 1999).
Many studies (\eg Mattila, Lehtinen \& Lemke 1999, Helou \ea 2000)
have confirmed the complex nature of spectral
energy distributions of disk galaxies in the $3 - 20 \umu$m range.
In addition to a continuum due to hot (or warm) dust there are bright 
IR-bands
at 3.3, 6.2, 7.7, 8.6, 11.3, and $12.7 \umu$m -- these are often called the
Unidentified Infrared Bands
(UIBs), due to the lack of understanding of their carriers.  These broadband
aromatic features are proposed to be the signature of Polycyclic Aromatic
Hydrocarbons (PAH; L\'eger \& Puget 1984).

The PAHs are an essential component in forming
the mid-infrared $[6.7/15]$ colour ratio which is emerging as a tracer
of star forming activity in galaxies (Vigroux \ea 1996, 1999, 
Sauvage \ea 1996, Dale \ea 2000, Roussel \ea 2001a, Helou 2000).  
The value $[6.7/15] \approx 1$ is expected in quiescent medium and PDRs,
while HII regions have $[6.7/15] < 0.5$ (\eg Cesarsky \ea 1996).
The $[6.7/15]$ ratio thus remains close to unity
for quiescent and mildly star forming galaxies, 
while it starts to drop for those with more vigorous star formation activity.
This mid-IR flux ratio has been also shown to correlate with the IRAS 
$[60/100]$ colour ratio, which is a well known indicator of activity level
in galaxies (Helou 2000, Dale \ea 2000, Vigroux \ea 1999).  
A two-component model of a galaxy as a linear combination of 
differing amounts of cold dust in cirrus clouds and warmer dust in HII 
regions has been seen as the explanation for the IRAS and IRAS-ISO 
colour-colour diagrams (Helou 1986, Dale \ea 1999, 2000).  
On the other hand, the
situation might be more complicated (see \eg Sauvage \& Thuan 1994), 
and for example it is possible that the proportion of star formation
in the disk relative to the central region of a galaxy plays a 
dominant role (\eg Vigroux \ea 1999, Roussel \ea 2001b). 

On another front, deep {\em ISO} galaxy counts 
(\eg Oliver \ea 1997, Taniguchi \ea 1997,
Elbaz \ea 1999a, Aussel \ea 1999, Flores \ea 1999, for ISOCAM counts)
have produced surprising results.  The
differential $15\umu$m counts show a remarkable upturn below
flux densities of 3 mJy and then a rapid convergence at approximately 0.4 
mJy.
This peak clearly requires strong (luminosity) evolution and can be
a result of strong mid-IR emission features, a new
population of sources, or some combination of these (Xu 2000,
Elbaz 1999b, Genzel \& Cesarsky 2000).
To understand these results and to develop a coherent picture of early
galaxy evolution, it is imperative to learn as much as possible about
the more local galaxies.

The ELAIS project (Rowan-Robinson \ea 1999, Oliver \ea 2000)
stands as a bridge between the very deep
galaxy surveys in the infrared mentioned above, and nearby galaxy
surveys (\eg Boselli \ea 1998, Dale \ea 2000, Roussel \ea 2001a).
ELAIS was the largest open time {\em ISO}-project with the driving
ambitious goal to study the un-obscured star formation out to
redshifts of $z \sim 1$.  Source counts in the mid-IR have been published 
in Serjeant et al.\ (2000)
and the far-IR counts in Efstathiou et al.\ (2000).

The aims of this paper are as follows:
In Section~\ref{obs} 
we present a subset of the ISOCAM ELAIS survey with near-IR
follow-up observations.  
%The near-IR $J$ and $K$ band
%survey itself was presented in V\"ais\"anen \ea (2000). 
A central new result of this paper, 
the calibration of the ELAIS data is performed in Section~\ref{calib} and
in the Appendix using the stars detected in our fields.
In Section~\ref{colcol} various near- to mid-IR colour-colour diagrams 
of the ELAIS galaxies are constructed and compared to
evolutionary models including the UIB features in the mid-IR, and to a 
local ISO galaxy sample.  
In Section~\ref{class} we attempt to
classify sources based on their NIR-MIR colours.  
Classifications such as this are expected to be helpful
in the future, eg.\ with {\em SIRTF} and {\em ASTRO-F} data,
when large numbers of galaxies with near-IR and mid-IR fluxes
become available without high-resolution spectra accompanying them at least 
in the first instance.  In Section~\ref{tracer} 
we discuss star formation properties of the
ELAIS galaxies, and the mid-IR and near-to-mid-IR colours as tracers of
star formation. Finally, active galaxies and 
extreme objects are discussed in Section~\ref{qsos}.

\section{Observations and Data}
\label{obs}

\subsection{ISO data}

The mid-IR ELAIS {\em ISO}-observations
were made with the ISOCAM LW2 ($6.7\umu$m)
and LW3 ($15\umu$m) filters,
covering ranges 5 -- 8.5 $\umu$m and 12 -- 18 $\umu$m, respectively.
For a description of the observations, data reduction,
and source extraction we refer the reader to Oliver \ea (2000) and
Serjeant et al. (2000).  At present the final reduction products are 
available only for the southern ELAIS fields (Lari \ea 2001) and thus
we use here the preliminary analysis v.1.3 ELAIS ISOCAM catalogue source 
list. This is somewhat deeper 
than the publicly available v.1.4 catalogue\footnote{Available at 
http://athena.ph.ic.ac.uk/elais/data.html} but otherwise equivalent (the 
latter is a subset of v.1.3).  At this stage the
detections are classified as `secure' (REL=2) or `likely' (REL=3). 
However, to have a reliable source list, {\em we will consider only
those detections with near-IR matches}, as discussed below.
Reliability and completeness in general of these versions of ELAIS catalogues 
will be discussed in more detail in Babbedge \& Rowan-Robinson (2002, in
preparation).  Part of our near-IR survey is in the N1 ELAIS
region, which was not observed at 6.7 $\umu$m.

\subsection{Near-Infrared data}
\label{nirdata}

The near-IR observations were carried out using the STELIRCam
instrument at the 1.2-m telescope of the F.\ L.\ Whipple Observatory
on Mount Hopkins.  A description of these $J$- and $K$-band
data (taken during 21 nights
between April 1997 and May 1999), reduction, as well as
photometry is found in V\"ais\"anen et al.\ (2000).  The survey area is
approximately 1 square degree, two thirds of it is in
the ELAIS N2 region (centered at RA=16h36m00s, 
DEC=$41\deg 06\arcmin 00\arcsec$)
and the rest in N1 (RA=16h09m00s, DEC=$54\deg 40\arcmin 00\arcsec$).
There is a small offset between the simultaneously observed FOVs in the 
$J$ and $K$ bands, resulting in slightly different source catalogues 
in the respective bands.

The 2MASS 2nd incremental data release (Cutri \ea 2000)
partially covers the N1 and N2 regions.
This allows us to directly cross-check our bright ($K < 14.5$) photometry
with 2MASS.  This is important also because we will later use 2MASS
data in connection with a comparion sample of nearby galaxies from the
literature.  The 'default' photometry of 2MASS was found to agree very well
with our photometry for both stars and galaxies.
Our data from Mt.Hopkins (while deeper due to
longer integration time) are, in fact, taken with a very similar telescope 
and instrument than the 2MASS data.

\subsection{Matching of mid- and near-IR data}
\label{matching}

\begin{figure*}
\centerline{\psfig{figure=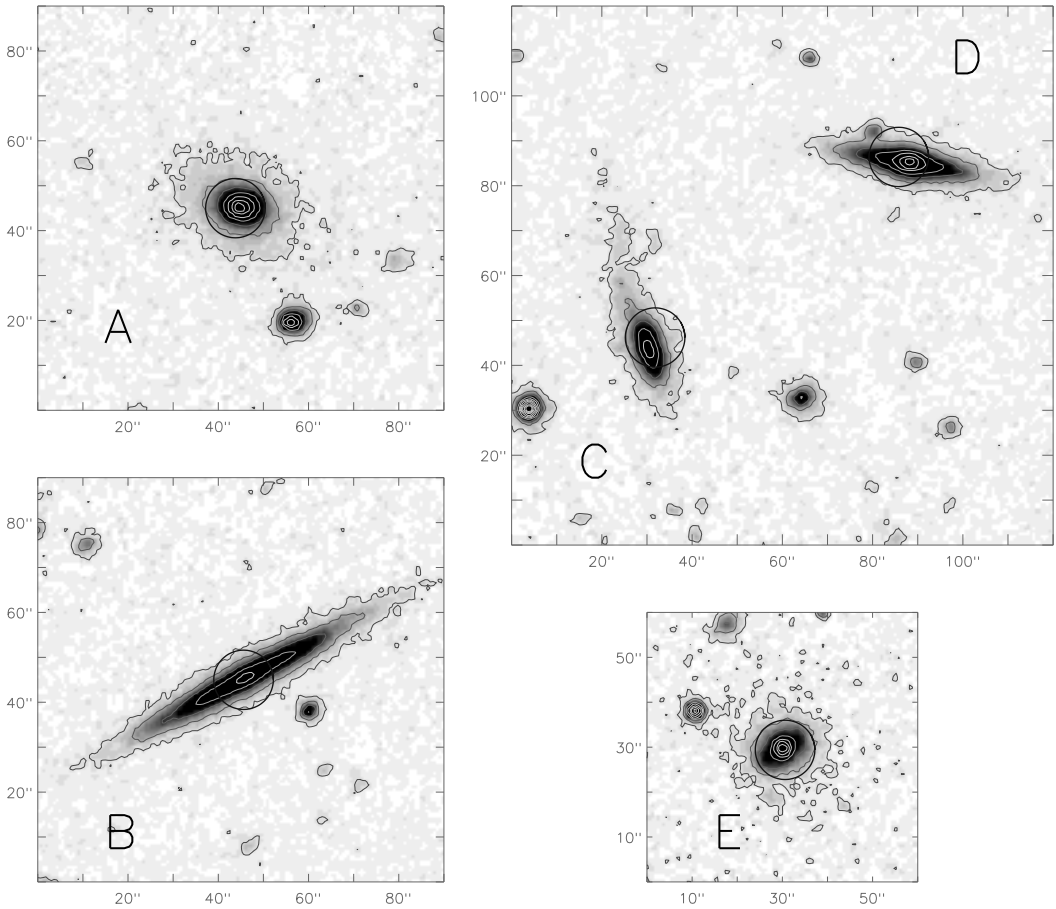,width=16cm}}
\caption{Examples of $K$-band images of
the brightest and largest (in NIR) galaxies in our sample.  
Positional error circles of 
13\arcsec\
diameter have been plotted around the ISOCAM detections.  The ISOCAM pixel
size is 6\arcsec.
Object `A' is an E/S0-type galaxy; the NED database
catalogues it with a name NPM1G +41.0441 and unknown redshift.
Object `B' has the largest extent of our sample. It is named UGC 10459,
lying at $z=0.03$ (NED), and it also is a radio-source
ELAISR20 J163507+405928 (Ciliegi et al.\ 1999).
Objects `C' and `D' form a galaxy pair, and `C' is the brightest mid-IR
source in our sample.  The pair's
redshifts or classifications are not available from the literature.
Object `E' is known as KUG 1632+414 ($z=0.03$) 
and it also is a radio and IRAS source.
NED classifies it as `spiral' and our near-IR image clearly shows a disk in
addition to a very bright unresolved nucleus.
The rest of the sources in our catalogue are much smaller.
}
\label{nir-early}
\end{figure*}

\begin{figure}
\centerline{\psfig{figure=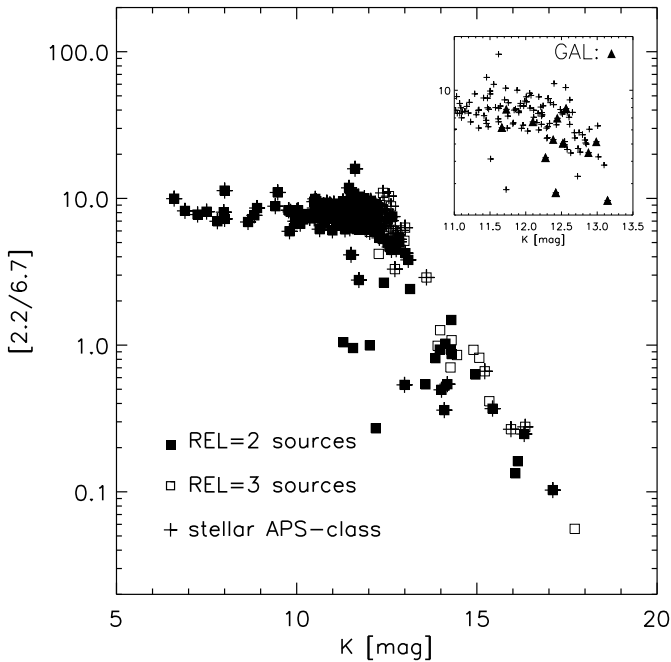,width=9.5cm}}
\caption{Near- to mid-infrared colour as a function of $K$-magnitude
($[2.2/6.7] \equiv f_{\nu}(2.2\umu{\rm m}) /
f_{\nu}(6.7\umu{\rm m})$.
Those objects which are classified (morphologically) as
stellar in the APS catalogue are overplotted with a cross.  All the 
brightest
objects are stars.  The inset shows a detail of the region where the stellar
population overlaps with galaxies (likely ellipticals).  
Galaxies are plotted as triangles and stars as crosses. 
}
\label{relstar_matches1}
\end{figure}

\begin{figure}
\centerline{\psfig{figure=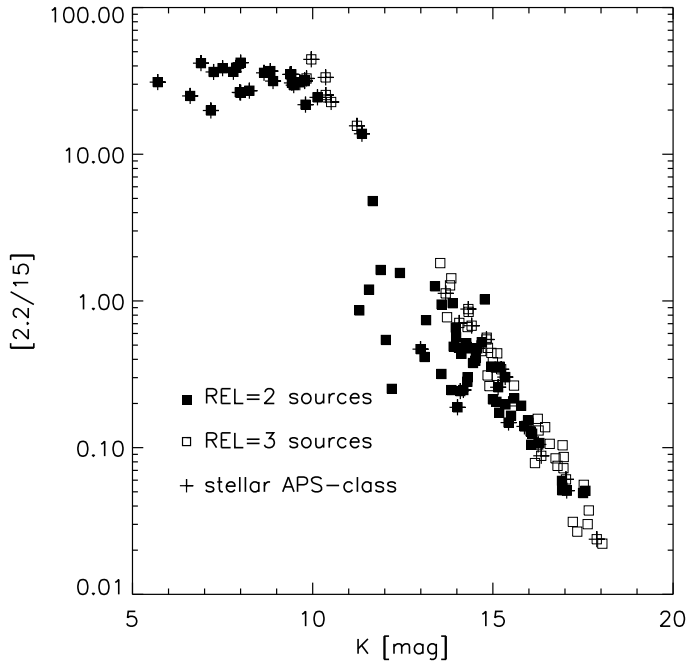,width=9.5cm}}
\caption{Same as the previous figure, but showing
the $K$ to $15 \umu$m colour.
}
\label{relstar_matches2}
\end{figure}

The ELAIS ISOCAM catalogue has 1322 and 2203 sources in total for {\em all}
ELAIS regions in the 6.7 and 15 $\umu$m bands, respectively.
These were matched with our
near-IR catalogue, which comes from a much smaller area.
The ELAIS v.1.3 catalogue includes
many double, or even multiple, detections from the edges of neighbouring
individual rasters and repeated observations --
thus we had to purge the catalogue.  We
searched for ISOCAM objects separated initially by 1\arcsec,
then 3\arcsec, and finally 6\arcsec\ -- at each step neighbours were merged
if they had the same near-IR counterpart.  
Ultimately, the matched and purged catalogue consists
of 217 and 158 NIR sources matched with the LW2 and LW3 ELAIS 
catalogue,
respectively.  Of these, 53 are common to both ISOCAM filters,
and due to LW2 coverage they are all in the N2 region. Table~1 gives the
total number of ISO sources, the NIR indentifications and their
classification per field, filter, and reliability class.
Notably, {\em all} those ISOCAM sources detected in both filters were 
identified, as well as all LW2 sources with REL=2. 

The probability of a chance appearance of eg.\ a $K=17$ mag object
within the 6\arcsec\ search radius is 0.03,
estimated from surface densities of near-IR objects
(eg.\ V\"ais\"anen et al. 2000).
More than 90 per cent of the matches are brighter than $K=16.5$ mag
-- we thus conclude that the purely positional matching is highly accurate.
And since we will be using only those mid-IR sources with a near-IR 
counterpart, we consider the sourcelist to be very reliable.

\begin{table*}
\begin{center}
\small
\caption{ISOCAM sources in the 1 sq.deg.\ near-IR survey area within the
ELAIS N1 and N2 fields, per band and reliability parameter; R(EL)=2
stands for a `secure' detection and R=3 for a `probable' detection. 
The columns give the total number of ISO sources, 
number of NIR identifications, and the classification of these matches (see
Section~\ref{stargal} for the classification).  Total numbers for 
each band are also shown. The 
row labeled LW2\&LW3 in N2 shows the numbers of sources with a 
detection in both 
ISOCAM bands. These objects were included in the respective LW2 and LW3 rows 
already; the breakdown by the REL parameter is not shown 
(however, for the galaxies these can be seen in table~2). }
\label{table0}
\begin{tabular}{llrrrr}
  &  & Detections & Identified & ~~~~~~Stars & 
	Galaxies \\
\hline
\multicolumn{2}{c}{\large N1} & & & \\
  LW3 & R=2 &  47  &  37  &   7  & 30 \\
      & R=3 &  59  &  12  &   1  & 11 \\
\hline
 \multicolumn{2}{c}{\large N2}   & & & & \\
  LW2 & R=2 & 170  & 170  & 141  & 29 \\
      & R=3 &  53  &  47  &  34  & 13 \\
  LW3 & R=2 &  68  &  60  &  19  & 41 \\
      & R=3 & 115  &  49  &   6  & 43 \\

 LW2\&LW3 & R=2\&3  &  53  &  53  &  24  &  29  \\
\hline
 \multicolumn{2}{c}{\large TOTALS}   & & & & \\
 LW2 &  &  223  &  217  &  175  &   42  \\
 LW3 &  &  289  &  158  &   33  &  125  \\
\hline
\end{tabular}
\end{center}
\end{table*}

The ELAIS fields were surveyed also with ISOPHOT at 90 and 175 $\umu$m.
Since the mid- to far-IR colours of ELAIS sources are discussed in another
work (Morel et al.\ 2002, in preparation), 
we do not discuss them further here, except to
note that 8 of our 29 galaxies with data from both ISOCAM bands and
near-IR photometry, also have 90 $\umu$m fluxes available.  
In addition,  20 of the 29  are included in the ELAIS VLA catalogue 
(Ciliegi \ea 1999).

\subsection{Photometry}
\label{photometry}

Since in this work we need to compare fluxes between nearby and distant
galaxies, total fluxes are required for both the near and mid-infrared.
In V\"ais\"anen et al. (2000) we found the `BEST' magnitudes
from SExtractor (Bertin \& Arnouts 1996) to be the most robust and accurate
over a wide range of magnitudes and
source profiles.  
The Kron-type `BEST'-magnitudes are presented in Table~2, but we
calculated also various aperture magnitudes and there is 
no difference in any final results if large enough apertures are used.
% (exceeding $\sim 10$\arcsec) magnitudes

The {\em ISO}-fluxes are measured from characteristic temporal
signatures of individual pixels, as described in Serjeant et al.\ (2000).
Instead of conventional aperture photometry the value of the peak pixel is
corrected to total flux using PSF modeling.  The adopted correction factors
were 1.54 at $6.7 \umu$m and 2.36 at $15 \umu$m.  The correction for the 
LW2 filter is more uncertain due to much undersampled PSF.  
Strictly, this correction is
appropriate for point sources only, which results in a potentially serious
underestimation of fluxes for extended objets.  However, the size of the
ISOCAM pixel is 6\arcsec, and the large majority of our sources are smaller
than this and we trust that the point source aperture correction 
gives an accurate value for them.  Nevertheless, we examined the 
largest ELAIS galaxies individually (using their NIR half-light radii and
testing with different apertures) to get an estimate of correction factors
to the mid-IR fluxes.
We conclude that only 4 of the galaxies, all of which are included in 
Fig.~\ref{nir-early}, definitely need a significant 
aperture correction.
For the largest galaxy in our sample
(ELAISC15 J163508+405933), referred to as `B'
in Table~2 and Fig.~\ref{nir-early}, we adopt fluxes from Morel et al 
(2002, in preparation),
modified in accordance with our new calibration (Section~\ref{calib}). 
The correction is very large, approximately a factor of 4.  
The other three galaxies labeled `A', `C', and `D',
respectively, are significantly smaller, and for these we adopt an approximate 
correction factor of 1.5.

\subsection{Star/galaxy classification}
\label{stargal}

We plot the NIR/MIR ELAIS data in Figs.~\ref{relstar_matches1}
and~\ref{relstar_matches2} as a function of $K$-magnitude,
using all the matched near-IR and ISOCAM detections.
We also matched all the sources in our field with optical data
from POSS plates
using the Automated Plate Scanner database (APS; Pennington \ea 1993),
which includes morphological star/galaxy separation.
Those objects with a stellar classification are identified as crosses
in the two plots.  

Stars clearly seem to lie in regions $[2.2/ 6.7] > 2$
and $[2.2/ 15] > 10$. We checked individually all objects in these
regions using our NIR data and were able to correct several 
ambiguous and erroneous APS-classifications (typically optically faint, red
objects). There are more stars in the $6.7 \umu$m matches, as expected,
and there is more overlap between separate populations in the 
$[2.2/6.7]$-plot. We verified that several galaxies, likely to be nearby 
ellipticals, lie in this overlap 
region which is shown in more detail in the inset of
Fig.~\ref{relstar_matches1}.
Henceforth, those objects which were morphologically
verified as stellar by near-IR and/or optical data, {\em and} have either
$[2.2/ 6.7] > 2$ or $[2.2/ 15] > 10$, are defined as stars.  The rest are
then classified as galaxies.

Several sources in the galaxies sample have a stellar APS classification.
While we did not attempt a comprehensive morphological classification of the 
faintest near-IR sources, many of them are obviously extended objects 
in the NIR data, and the erroneous
APS classification is just due to faintness of objects.  Some of them are 
however point-like also in our data,
and thus are potentially interesting cases, to which we
will return in Section~\ref{qsos}.  However, this group might still
include rare dust-shell stars.

It is interesting to note the proportions of stars and galaxies 
(see Table~1): in the 
near- and mid-IR matched catalogue 81 per cent of the 6.7 $\umu$m sources 
are stars.  At 15 $\umu$m only 21 per cent of the objects are stars.

\subsection{Flux calibration of ISOCAM ELAIS data using stars} 
\label{calib}

Before removing the stars from further consideration in this paper, we use 
them for an accurate flux calibration of the ELAIS {\em ISO}-data.
We match observed near-IR and mid-IR colours to corresponding model colours 
of infrared standard stars, and are able to derive the flux calibration
for the ELAIS ISOCAM data with better accuracy than done previously.
The derivation is performed in the Appendix~A: we adopt values of
1.23 and 1.05 ADU/gain/s/mJy for the LW2 and LW3 filters, respectively,
\ie the catalogue v.1.3 values for LW2 and LW3 have to be multiplied 
by these factors to have fluxes in mJy.  Note that the factors were not 
included in Figs.~\ref{relstar_matches1} and~\ref{relstar_matches2} above.
(The LW3 calibration is in disagreement with the one performed
in Serjeant et al. (2000), where a value of 1.75 ADU/gain/s/mJy was found.)
Our values are in good agreement with the ISOCAM handbook values of
2.32 and 1.96 ADU/gain/s/mJy (Blommaert 1998), where an additional factor
of 2 correction for signal stabilization has been included (see Appendix for
details).  This lends strong support for the accuracy of the reduction and
photometric techniques used in the creation of the ELAIS Preliminary 
catalogue.

Furthermore, we can use the bright stars to estimate the completeness
of the ELAIS ISOCAM catalogue.  Using the mean MIR/NIR flux ratio for
stars (see Appendix and also Figs.~\ref{relstar_matches1}
and~\ref{relstar_matches2}) we calculate the expected mid-IR fluxes of all 
bright near-IR stars in our field, and then check whether they actually are
included in the ELAIS catalogue.  The results are as follows: the 15 $\umu$m
catalogue is essentially complete above 2 mJy. Below this 
the completeness begins to drop rapidly.  
Six stars out of 20 between 1.0 and 2.0 mJy are detected.  The
6.7 $\umu$m band is essentially complete above 1.5 mJy. 
The completeness above 1 mJy is 80 per cent, while only 18 out
of 73 stars between 0.5 and 1.0 mJy are detected.  These numbers are 
consistent with those derived by Serjeant et al.\ (2000) who
find the 50 per cent completeness limits
for the whole ELAIS survey to be $\sim1$ mJy for both bands (using the
calibration of this paper for the 15 $\mu$m data).

\begin{figure}
\centerline{\psfig{figure=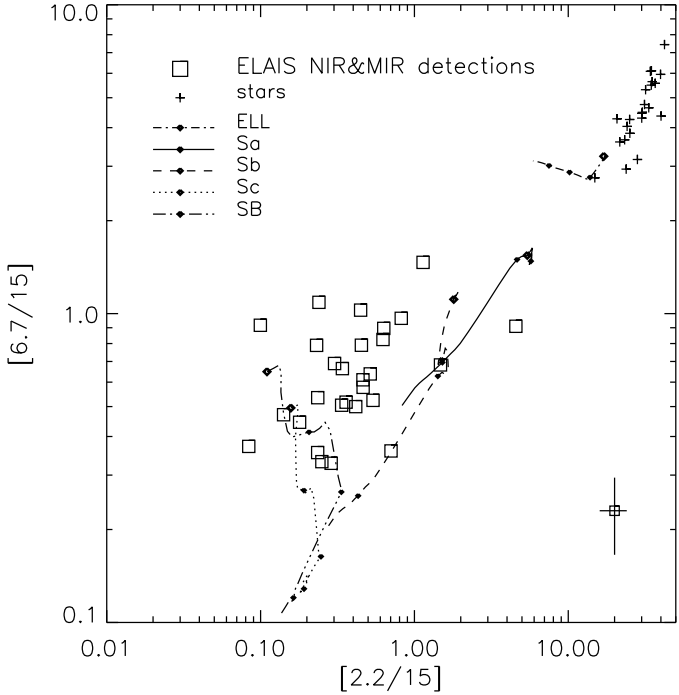,width=10cm}}
\caption{The $[6.7/15]$ mid-IR colour ratio vs.\ $[2.2/15]$ colour of all 
ELAIS sources in our fields.  Those objects which were previously
defined as stars are shown as crosses.
GRASIL model predictions for galaxies are overplotted.
The model colours are for a range of $0 < z < 1.0$, with the largest solid
symbol marking $z=0$ and the others the positions for $z=0.25, 0.5$, and 
0.75.  Average, conservative error bars are shown at the lower right corner.}
\label{6715-vs-k15-ii}
\end{figure}

\begin{figure}
\centerline{\psfig{figure=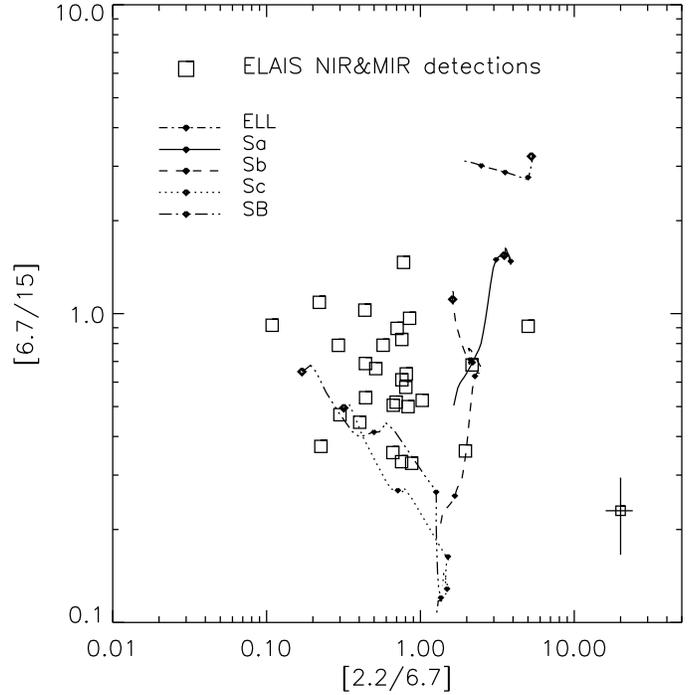,width=10cm}}
\caption{The $[6.7/15]$ ratio vs.\ $[2.2/6.7]$. Typical error is again
indicated at lower right corner.}
\label{6715-vs-k67}
\end{figure}

\section{Colour-colour distributions}
\label{colcol}

\subsection{Models}
\label{models}

The detailed modeling of NIR/MIR colour-colour distributions and resulting
interpretations of physical properties and star formation rates
of the galaxies have to wait for more comprehensive spectral and
redshift data.   However, it is very informative to check
the expected redshift effects on the colour-colour diagrams.
We chose to use a set of models from the GRASIL
code\footnote{Libraries of selected models are publicly available at
http://grana.pd.astro.it/grasil/modlib/modlib.html} (Silva \ea 1998).
We used four evolving GRASIL SEDs (elliptical, Sa, Sb, and Sc) to compute
K- and evolutionary corrections and observed colours as function of
redshift, for our $J$- and $K$-bands as well as the ISOCAM filters.
Appropriate filter curves were used to convolve the SEDs, and a
colour-correction was performed in case of ISOCAM to
be consistent with the convention of measurements (see Blommaert 1998).
These four models are used in the following to compare with the data.
In addition, to represent a starburst, we simply took the SED of an
early-type spiral at the age of 2 Gyr, and held it constant at all
epochs.  In the wavelength range considered here the SED is similar to that
of M82 and also not too unlike Sc's in general
(see Silva \ea 1998, Schmitt \ea 1997).
The cosmology used in the plots is $q_{0} = 0.15$ and $H_{0} = 50 {\rm km}
{\rm s}^{-1} {\rm Mpc}^{-1}$, though changing this does not have significant 
effects in the redshift ranges considered.
It should also be kept in mind, that
the models represent total luminosity of a galaxy, while the observations
in practice are performed with a given aperture (although `total flux' in
photometry is attempted, as discussed above).

\subsection{NIR-MIR colours of ELAIS galaxies}
\label{colours}

In general, the emission from galaxies in near-IR
bands is due to the stellar contribution,
the $6.7\umu$m carries information on the PAH contribution, and
any strong $15 \umu$m emission would indicate warm dust.
There are thus several colour indices which may be useful in studying
the relative strengths of these components and processes.  For example,
the $6.7/15 \umu$m flux ratio is expected to 
trace activity in the ISM of galaxies.

\begin{figure*}
\centerline{\psfig{figure=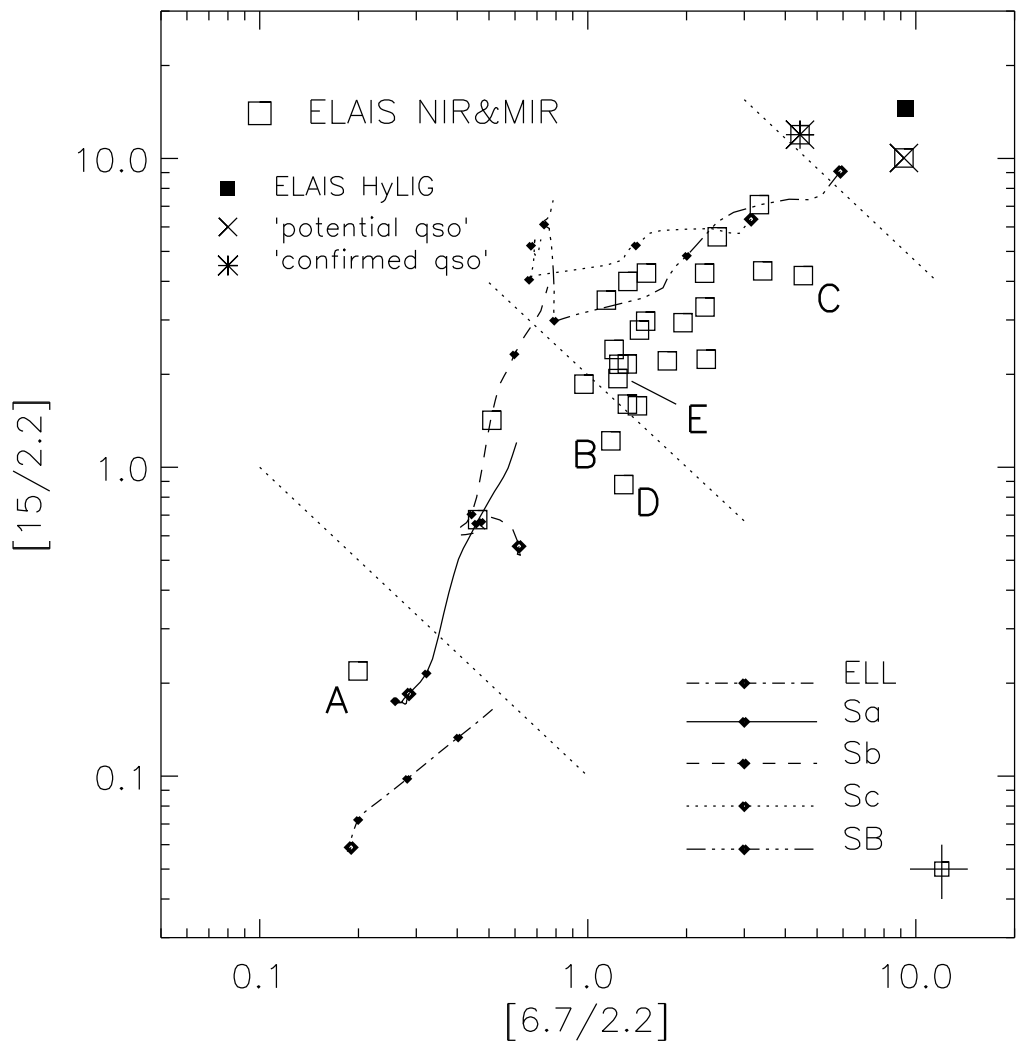,width=15cm}}
\caption{The 6.7 and 15 $\umu$m fluxes normalized with the $K$-band
flux, i.e.\ by the stellar contribution to the brightness of the galaxy.
The strengths of the mid-IR fluxes are seen to correlate strongly, and
the difference in the relative strength of mid-IR flux ranges nearly
two orders of magnitude. The GRASIL models are overplotted again.
The bright galaxies of Fig.~\ref{nir-early} are labeled from A to E.
The dotted lines roughly separate areas for different types of galaxies --
see Section~\ref{class}.  In addition, we have overplotted a hyperluminous
infrared galaxy ($z=1.1$) detected in another ELAIS field (see Morel \ea 
2001)
and two `potential quasars' discussed in Section~\ref{qsos}, one of which is
a confirmed QSO at $z=1.14$. Typical error is at bottom right corner.}
\label{15k-vs-7k}
\end{figure*}

Figs.~\ref{6715-vs-k15-ii} and~\ref{6715-vs-k67} show the $[6.7/15]$ 
ratio against $[2.2/15]$ and $[6.7/15]$.  The first compares the relative 
strength of the stellar and warm ISM component.  
Objects to the right are dominated
by stellar emission, and those to the left by warm dust, while the vertical
axis tells about the heating activity and the relations of PAHs and warm 
dust.  The second figure depicts the stellar vs.\ PAH contribution.
To further study the $[6.7/15]$ ratio, the ISOCAM
bands are plotted against each other in Fig.~\ref{15k-vs-7k},
normalizing with the NIR flux, which in addition to stellar light is 
expected to be a good measure of stellar mass in a galaxy
(\eg Kauffmann \& Charlot 1998).
The implications of this will be discussed more in Section~\ref{discussion}.

Models presented in
Section~\ref{models} are overplotted in all the colour-colour figures
for a range $z = 0-1$.
Note that the UIB
features move rapidly beyond the $6.7 \umu$m filter with redshift, which
results in decreasing $[6.7/15]$ in models including strong PAH emission
(especially Sc and starbursts).
The colour of Sa type galaxies, on the other hand, starts
to change only at $z\sim0.75$.  Ellipticals at zero-redshift
occupy the same region as red stars, as expected.

Most of the ELAIS galaxies with data in both mid-IR bands
appear to group at a region where the models
predict low-redshift, $z=0.1 - 0.4$, late-type Sc spirals.  According to
the models, the near- to mid-IR SEDs of all spirals would look fairly 
similar at $z\sim1$.  However, it is unlikely that such
objects are detected to the ELAIS survey limits.  AGN on the other hand
are expected to lie
at the extreme upper right in Fig.~\ref{15k-vs-7k} due to their steeply 
rising continuum (\eg Laurent \ea 2000).

\begin{figure}
\centerline{\psfig{figure=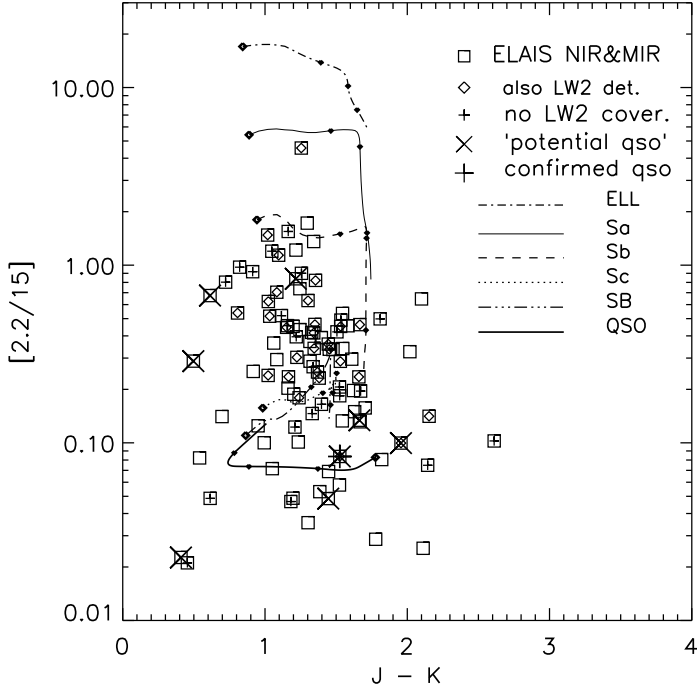,width=10cm}}
\caption{The $[2.2/15]$ ratio
plotted as a function of $J-K$ colour.  For most normal galaxies
$J-K$ is expected to be between 1 and 2. 
Those sources which are detected in both
mid-IR bands are marked with diamond, and those which do not have 
LW2 coverage are marked with a cross.
Most of the sources missed by the $6.7 \umu$m survey (\ie empty squares)
have a very small $[2.2/15]$ ratio, expected from moderate redshift Sc and 
starburts galaxies as predicted by GRASIL. See Fig.~\ref{derseds} 
for more details.
One confirmed QSO ($z=1.142$) is marked with a large asterisk, and 
`potential
QSOs' discussed in Section~\ref{qsos} with large diagonal crosses.
The QSO-model (from Schmitt \ea 1998) curve is marked at half-redshift
intervals, while other models are with 0.25-intervals, as before.
}
\label{j-k_plot}
\end{figure}

The two mid-IR filters detect surprisingly different populations.
As can be seen from Table~1, of the 97 identified galaxies 
which are from an area covered with both mid-IR bands,
only 29 are common to both LW2 and LW3.  There are
55 galaxies detected only at $15 \umu$m, and 13 galaxies detected only at
$6.7 \umu$m.
For those ISOCAM sources with a detection in only one mid-IR band 
the $J-K$ colour might provide additional clues. For example, 
starbursting galaxies should have very red $J-K$ 
colours.   Fig.~\ref{j-k_plot} plots the $[2.2/15]$ against $J-K$. 
While there are a number of sources with a red $J-K$,
they do not constitute a large population. 
More strikingly, compared to Figs.~\ref{6715-vs-k15-ii} and~\ref{15k-vs-7k},
there are many more sources at a low $[2.2/15]$ ratio.

\begin{figure}
\centerline{\psfig{figure=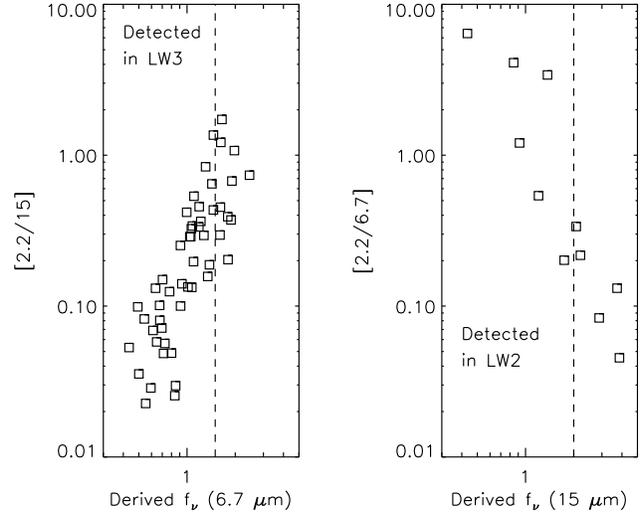,width=9cm}}
\caption{Galaxies detected in only one ISOCAM filter. Fluxes in the other
band are derived from their expected
correlation to NIR/MIR ratio using models, ELAIS galaxies,
and local galaxies discussed in Section~\ref{compsamp}. 
The dashed vertical lines show the expected completeness levels.
Those detected at $15 \umu$m but missed in the
$6.7\umu$m band are plotted on the left.  According to $[2.2/15]$ 
these galaxies should typically be Sc's, starbursts,
and perhaps AGN. 
Galaxies detected in $6.7 \umu$m but missed in the
$15\umu$m band are on the right. Many of the missed
galaxies are early types with high $[2.2/6.7]$, though there are also 
QSOs included.
}
\label{derseds}
\end{figure}

While the most extreme
sources are too faint to acquire any definite morphological information
from our data, we can constrain the nature of the sources missed in the
6.7 $\umu$m band by examining detections limits.  
Since the 15 $\umu$m fluxes of the missed 51 galaxies range from 1 to 3 mJy,
typical $[6.7/15]$ ratios should be around 0.45 to be consistent with the
80 per cent LW2 completeness limit detection limit.
This implies Sc galaxies or starbursts around $z\approx 0.15$
or Sb's at $z\sim 0.5$.  Accordingly, most $[2.2/15]$ ratios of the missed 
sources (empty squares in Fig.~\ref{j-k_plot}) do lie by the Sc and starburst
model curves.  We also derived a rough estimate for the
expected 6.7 $\umu$m flux from a mean correlation of $[15/2.2]$ with 
$[6.7/2.2]$, using the ELAIS sources in Fig.~\ref{15k-vs-7k} and also a 
comparison sample discussed below in Section~\ref{compsamp}.  The derived
$f_{\nu}$(6.7$\umu$m) are shown in Fig.~\ref{derseds}. 
Galaxies with the lowest $[2.2/15]$ ratios fall below the LW2 detection 
limit while still being detected in LW3.  It is thus clear that 
the faintest late type spirals and starbursts 
make up the majority of LW2-missed sources.
However, statistically we should find only approximately 5
sources with LW2 fluxes between 1 and 2 mJy in the figure. 
There are around 20, many of them with a $[2.2/15]$ ratio
typical of earlier type spirals (Sb's).  These might harbour some form of 
activity resulting in a lower than expected $[6.7/15]$ ratio.

\begin{figure}
\centerline{\psfig{figure=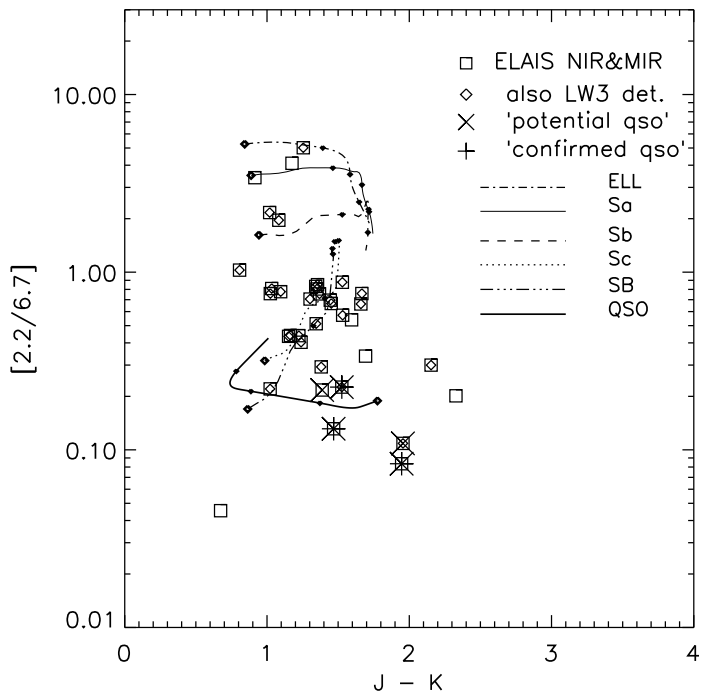,width=10cm}}
\caption{The $[2.2/6.7]$ ratio plotted against $J-K$ colour. Some of
the objects missed by LW3 seem to be nearby early-type galaxies.  
In fact, there are several more such cases, but they lack coverage in
one or the other NIR band. 
Other LW3-missed objects have low $[2.2/6.7]$ and lie close to the
QSO-model curve.
`Potential QSOs' discussed later in Section~\ref{qsos}, are marked with
large diagonal crosses;  three of these are actually catalogued quasars
and are overplotted with a large cross (only one of them is detected by LW3).
}
\label{j-k_plot2}
\end{figure}

The objects seen in $6.7\umu$m but not in $15 \umu$m are 
plotted in Fig.~\ref{j-k_plot2}.
As seen there and in Fig.~\ref{derseds}, 
almost half of the LW3-missed objects appear to be
ISM-deficient early types.  However, since the LW3 catalogue
should be more than 90 per cent complete above 2 mJy,
the derived $f_{\nu}$($15\umu$m) at least for some of the galaxies might
be too high.  Surprisingly, two confirmed QSOs are among LW3-missed 
objects (see Section~\ref{qsos}). However, numbers are small, 
and the 6.7 $\umu$m fluxes are low, close to the detection limit, 
so it is difficult to conclude anything definite.

\subsection{Comparison sample}
\label{compsamp}

In order to compare our resulting ELAIS
near- to mid-IR colours to a local sample of galaxies observed with {\em 
ISO},
and to discuss how well the galaxy types can be separated with near- and 
mid-IR
colours, we made use of the data-sets of Roussel et al. (2001a), 
Dale et al. (2000), and Boselli et al. (1998).  Naturally, there exists
a large body of work performed with {\em IRAS} galaxies establishing
near- and mid-IR databases (\eg Spinoglio \ea 1995) -- however, to avoid
complications of band conversions we restrict ourselves only to recent
{\em ISO}-data.
The Roussel et al.\ set consists of nearby spirals, and it
includes a subset of the Boselli sample, which are Virgo cluster galaxies.
The Dale et al.\ sample are galaxies from the {\em ISO}
U.S. Key Project `Normal
Galaxies'.

The main difficulty in the comparison are the various photometric techniques
used both in the near-IR and {\em ISO} data (see \eg Spinoglio \ea 1995).
A large number of the nearby galaxies have near-IR data available from NED.
However, to have consistent photometry
we decided to only
use those galaxies for which there were 2MASS data available from the
2nd incremental data release.
The 2MASS catalogue lists numerous magnitudes of which we used
the default `fiducial circular magnitudes at 20 mag/sq.arcsec',
because of the wide range of sizes and shapes of
the galaxies in the comparison sample.

Boselli et al.\ (1998) give their own near-IR photometry.
We performed a cross-check of their $K$-band photometry between 2MASS values
and found a fairly good consistency: apart from some Sc-type galaxies
2MASS isophotal magnitudes were 0.15 mag fainter than the Boselli
values, with a scatter of 0.31 mag. 
That the level of Boselli and 2MASS magnitudes are so close, and
that the sizes of galaxies in the other two samples are not significantly
different from the Boselli galaxies, 
gives us confidence to use 2MASS magnitudes for the nearby 
comparison galaxies,
and to directly compare them to the `total magnitudes' of the ELAIS 
objects.  Nevertheless, for consistency reasons with Dale and Roussel 
samples, we decided to use only those galaxies
from Boselli \ea which had 2MASS photometry available.
In the following, with the Dale and Roussel samples we mean those galaxies
from the respective original works for which we found 2MASS magnitudes.
Our Boselli sample means those galaxies from Boselli \ea with
2MASS photometry, and those galaxies excluded, which already are included in
the Roussel sample.

As for the {\em ISO}-data, the standard CAM
Interactive Analysis (CIA) packages were used for pre-processing of raw
data in all of the comparison data here, as well as
the ELAIS data.
The {\em ISO}-fluxes of all the comparison samples are
calculated from {\em maps} resulting from the application of
CIA/IDL procedures (see Roussel et al. 2001a for a detailed description
of the reduction process).
Fluxes in all these comparison samples were calculated using 
various apertures, and all claim their 
fluxes to represent total values at better than 20--30 per cent accuracy.
There is a systematic difference by approximately a factor of 1.4 in fluxes
of some galaxies common to the Roussel \ea and Dale \ea catalogues (see 
Roussel \ea 2001a), none of these are however included in our subsamples.
The mid-IR fluxes in Boselli \ea (1998) are not published apart from SED 
plots; the values were provided by A.\ Boselli.

As mentioned earlier, in contrast to the photometry of these 
comparison samples,
the mid-IR ELAIS fluxes are values derived from peaks in time histories
of individual pixels and corrected for PSF effects.
However, we are confident that the ELAIS ISOCAM
fluxes are close to the true total values 
(see Sections~\ref{photometry} and~\ref{calib} and the Appendix~\ref{appdx}).

\begin{figure*}
\centerline{\psfig{figure=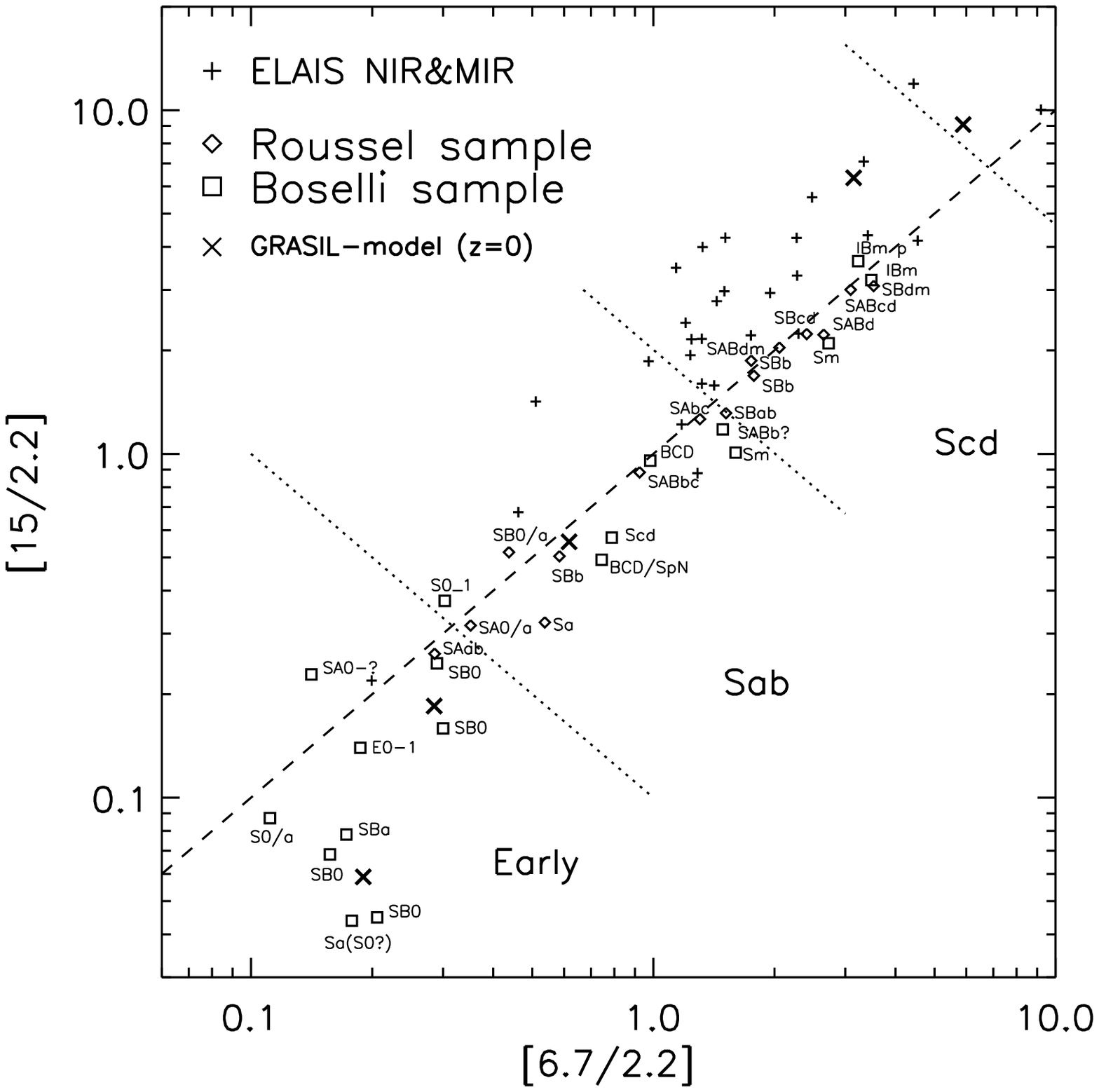,width=15cm}}
\caption{ The 6.7 and 15 $\umu$m fluxes from Roussel
et al. (2001) and Boselli et al. (1998), normalized with the $K$-band
fluxes.
Corresponding near-IR data is from 2MASS.
The ELAIS data are overplotted as small crosses and
the GRASIL models are shown for $z=0$ only (see Fig.~\ref{15k-vs-7k} for
redshift dependence).  The dashed line shows the
one-to-one ratio of 15 and 6.7 $\umu$m fluxes.  We have separated regions
roughly corresponding to different types of galaxies with dotted lines --
see text for details.
}
\label{ROUBOS-15k-vs-7k}
\end{figure*}

\section{Discussion}
\label{discussion}

\subsection{Classifying ELAIS galaxies}
\label{class}

Unfortunately at present we do not have optical imaging to definitely 
classify our ELAIS galaxies.  Classification with the help of GRASIL models 
was briefly discussed along with NIR/MIR colours in Section~\ref{colours}.
How accurate can the classification of normal 
galaxies be using near and mid-IR colours?  For example,
Sauvage \& Thuan (1994) find only loose correlations between IRAS colours
and morphological type in their large sample.
There would be interest in having a NIR/MIR photometric classification,
in anticipation of forth-coming very large galaxy surveys by \eg SIRTF.
Here we will investigate whether the comparison sample 
presented above has correlations between morphology and NIR/MIR
properties and see what these would imply for our ELAIS galaxies.

Figure~\ref{ROUBOS-15k-vs-7k} shows the Roussel and Boselli samples 
in a MIR/NIR two-colour diagram along with ELAIS sources.
There is clear trend: the
type of galaxy becomes systematically
later upward along the diagonal.  This should not be a great surprise,
since essentially the progression
shows the overall amount of emission from the ISM in the galaxy 
increasing.
The average $[15/2.2]$ for early types (Sa and earlier) is 0.2 whereas for
the late types  (Sc and later) $[15/2.2] \approx 2.2$.  

\begin{figure*}
\centerline{\psfig{figure=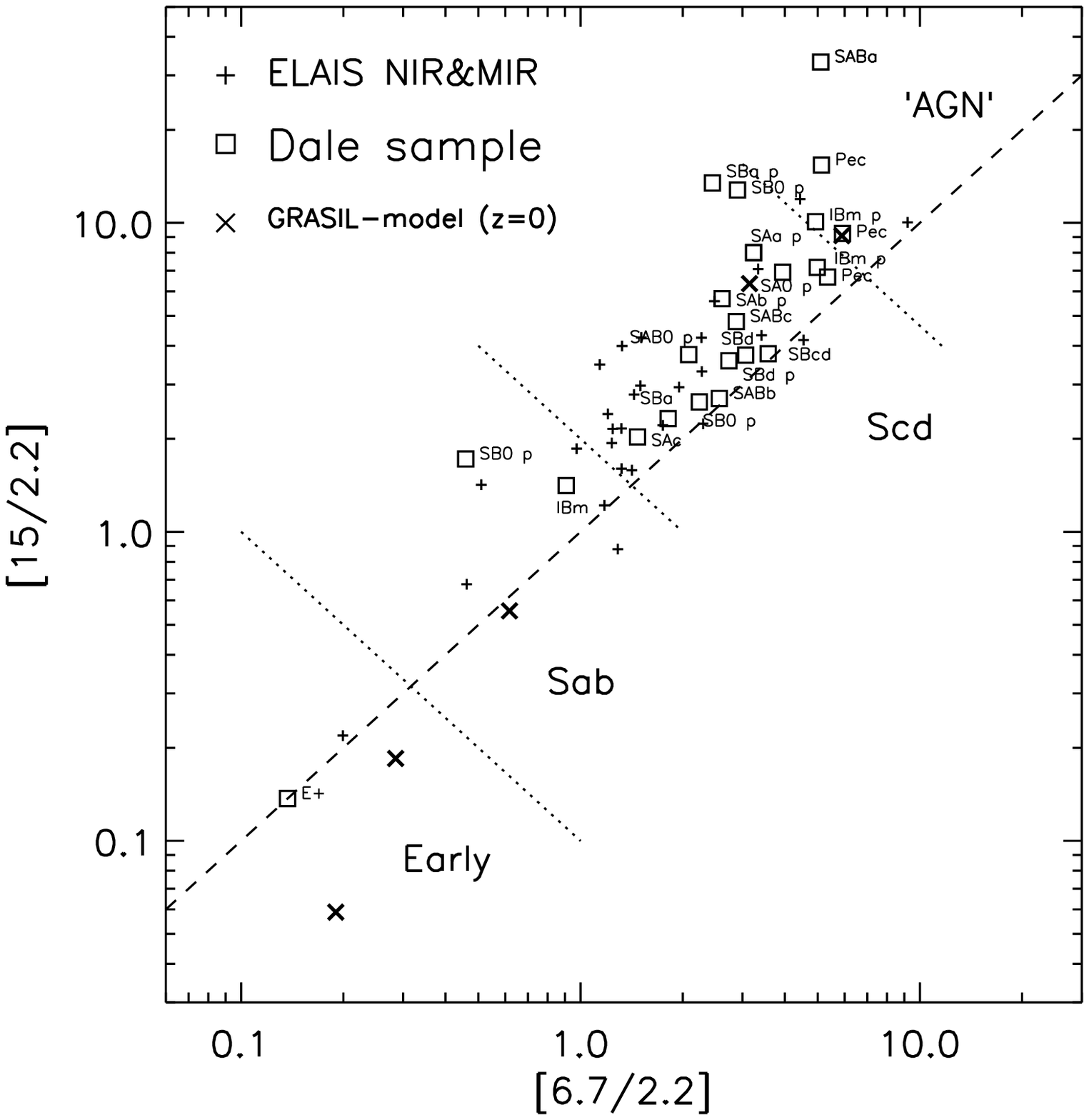,width=15cm}}
\caption{The 6.7 and 15 $\umu$m fluxes from Dale et al. (2000)
normalized with the $K$-band flux.  Near-IR data is from 2MASS, except
for the galaxy at lower left region (NGC 6958, NED magnitude;
included merely to have one example of an elliptical).
The region loosely labeled `AGN' is expected to include sources 
such as QSOs, ULIRGs, and strong starbursts.
ELAIS data, GRASIL models, and dashed and dotted lines 
are presented as in the previous figure.}
\label{DAL-15k-vs-7k}
\end{figure*}

Figure~\ref{DAL-15k-vs-7k} shows the Dale sample in the same fashion.
The sample covers a wide range of morphological types,
but it is evident that barred galaxies are plentiful and especially 
galaxies
which have been attached with a peculiar (`p') morphology in
addition to a regular Hubble type.
The trend seen in the Roussel and Boselli samples is not clear at 
all.  The whole sample groups strongly 
towards the Sc-model colour, including the morphological early type
galaxies.  The peculiars ('p' and purely `Pec') and irregulars tend to have
highest MIR/NIR ratios.  

The Dale sample also has many more
galaxies with significantly lower $[6.7/15]$ compared to the 
Roussel/Boselli  sample.  The difference mainly comes from galaxies 
with high MIR/NIR ratios.  This is true regardless of the $\sim 40$ 
per cent discrepancy in the photometry  mentioned earlier. 
The Boselli and Roussel galaxies on the other hand
are strongly concentrated along the one-to-one correlation line where 
$[6.7/15] \approx 1$, where the galaxies are supposedly dominated by
quiescent ISM.  We will return to this point in Section~\ref{clues}. 

As can be seen in Figs.~\ref{ROUBOS-15k-vs-7k} and~\ref{DAL-15k-vs-7k}, 
we have divided the diagram into four regions with the dotted lines: 
the areas
roughly correspond to low-redshift early-types, Sab spirals, Scd's, 
and 'AGN' (the latter class includes several types of active 
sources \eg  QSOs, Seyfert nuclei, ULIRGs, strong starbursts). 
The dividing lines in the figures can be obtained from $\log [15/2.2] = 
-\log [6.7/2.2] + b$ where is $b= -1.0, 0.3$, and 1.67 starting from 
the lower left, respectively.
The Roussel and Boselli galaxies fall very well into their areas.
Disregarding BCDs, there are only 6 galaxies out of 34 in a `wrong' area,
and of these, 5 are very close to the border-lines.
This classification does not work as well
for the Dale sample though.   We conclude that according to the nearby 
comparison sample the NIR/MIR two-colour 
diagrams do discriminate between types of normal galaxies,
especially those which have $[6.7/15] \approx 1$.

Where are our ELAIS galaxies in this classification?  
The ELAIS sample as a whole clearly groups towards the late Hubble types.
However, as seen above, NIR/MIR flux
ratio may not be a good indication of morphological type for those 
galaxies with low $[6.7/15]$.  Nevertheless, of the 29 galaxies in 
Table~2 (see Fig.~\ref{15k-vs-7k}) there are 
21 galaxies in the Scd-region and 5 in the Sab-region.  Two are found
in the upper-most region in the far right -- in Section~\ref{qsos} they 
are shown to be potential AGN.  Only one galaxy 
seems to be an early type, though it does have excess $15 \umu$m flux.
Indeed, early-type galaxies have been shown to have widely differing 
amounts of dust (see Madden, Vigroux \& Sauvage 1999, and 
references therein).  `Traditional' ellipticals, with no significant 
ISM presence, would not have 
been seen at all by the LW3 filter in the ELAIS survey. 
As shown in the inset 
of Fig.~\ref{relstar_matches1}, there are several probable ellipticals
which are detected only in LW2.

All the largest galaxies which show clear morphology in our data 
(Fig.~\ref{nir-early}, \ie those labeled in Fig.~\ref{15k-vs-7k}) 
are in consistent classification areas: 'A' is the early type galaxy 
and the rest are spirals.
Object E is a disk galaxy with a very bright compact 
nucleus. It has the lowest $[6.7/15]$ 
ratio of these five bright galaxies, indicating star-formation, as will 
be discussed next.

\subsection{Tracing star-formation activity}
\label{tracer}

\subsubsection{Star-formation tracers}

Much discussed tracers of star-formation include the
H$\alpha$ emission of a galaxy, the UV-continuum, and total
far-IR luminosity.  It is also well-known that star-formation in galaxies
occurs in two very distinct places: in the disks of spirals and
in compact circumnuclear regions (for a comprehensive review,
see Kennicutt 1998).
In principle the mid-IR could help in solving some of the
uncertainties related to the mentioned diagnostics: eg.\ mid-IR is
certainly less prone to extinction
than UV and H$\alpha$ studies; it could also help in determining the
heating source of IR emission, which affects the accuracy of the FIR
diagnostic.  The FIR tracer is known to work well for circumnuclear 
starbursts
-- it is in the disks of normal galaxies, where help would be needed.

If the mid-IR is to be useful as a star-formation rate (SFR) indicator,
calibrators with the other methods are necessary due to the complexity of
theoretically deriving an accurate relation between mid-IR emission and
amount of young stars.  Indeed,
H$\alpha$ emission has been shown to correlate with mid-IR luminosity
{\em in the disks} of spiral galaxies (Roussel et al. 2001b,
Vigroux et al. 1999, Cesarsky \& Sauvage 1999).  Also far-IR seems to
correlate linearly with mid-IR and H$\alpha$ if only disks are considered,
and SFR can thus be estimated (Vigroux et al 1999, Roussel \ea 2001b).
The relations do not hold in regions of more intense
star-formation (eg.\ nucleus), and thus nuclear
star-formation could confuse a global SFR determination.  Vigroux et al.\
and Roussel \ea argue that this precisely is the reason for
non-linearity in global far-IR vs.\ H$\alpha$ relations.
Thus, only limits to SFR can be calculated
from integrated mid-IR luminosity, and the need for information
on the proportions of disk and nuclear IR-emission is high-lighted.

\subsubsection{Clues from $f_{\nu}(6.7\umu{\rm m}) / 
f_{\nu}(15\umu{\rm m})$ and near-IR to mid-IR}
\label{clues}

\begin{figure*}
\centerline{\psfig{figure=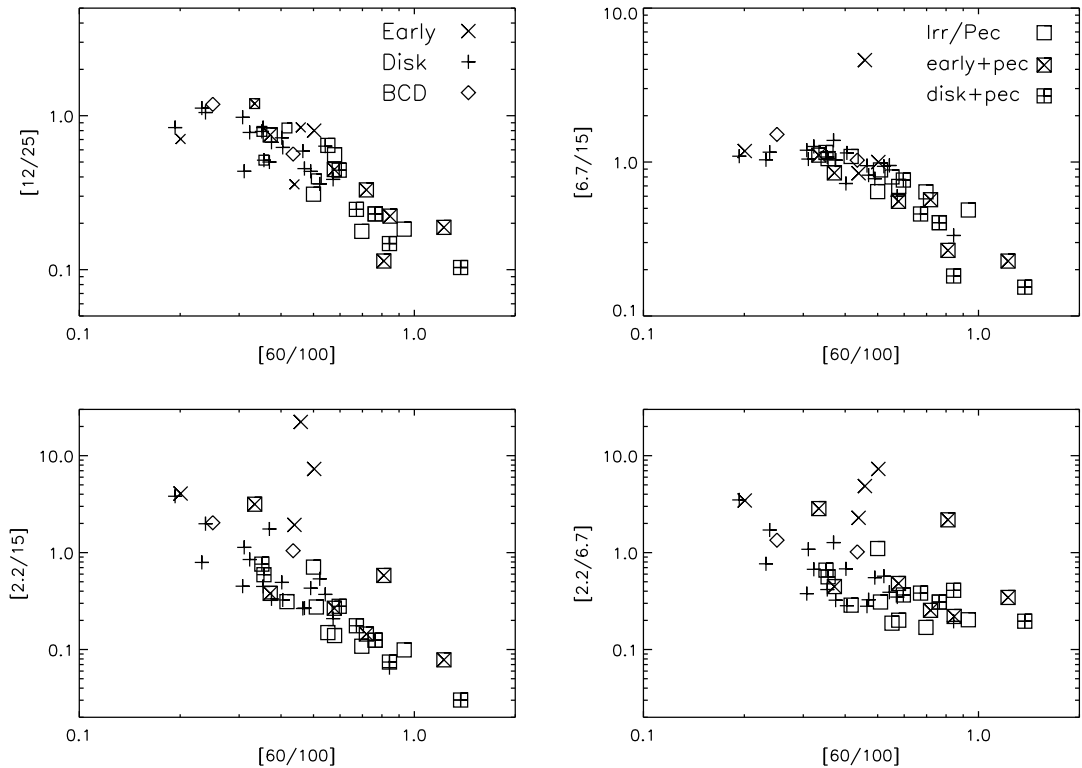,width=15cm}}
\caption{Comparing the {\em IRAS} $[60/100]$ colour and star-formation
tracer to near- and mid-IR colours of all the comparison sample galaxies
with available IRAS fluxes.  Early types (until Sa), disk galaxies, 
irregulars, and BCDs are plotted with different symbols. Those classified
as peculiar are overplotted with squares.
Note that most of the early-type Boselli galaxies
had only upper limits in the {\em IRAS} data, and are not plotted (they 
would populate the relatively empty region at $[6.7/15] > 2$ in the 
upper right panel).  
}
\label{fircols}
\end{figure*}

The mid-IR flux ratio is helpful in tracing the star-forming activity,
as discussed before.  We also expect the near-IR to add to the 
information.
To illustrate these effects, Fig.~\ref{fircols} shows all those
galaxies from the comparison sample discussed
in Sections~\ref{compsamp} and~\ref{class}, which had {\em IRAS}
fluxes available.
The upper left panel shows an {\em IRAS}-colour diagram and upper right
the {\em ISO-IRAS} colour distribution.  The lower panels plot the
NIR/MIR colours against the IRAS $[60/100]$ colour. The galaxies are
differentiated by morphology to ellipticals/lenticulars, disks, and
irregulars/peculiars -- a galaxy may have both a lenticular or disk and
a peculiar classification.

The previously observed trend (Helou 1999, 2000, Dale \ea 2000, 
Vigroux \ea 1999), that $[6.7/15]$ first remains
fairly constant and starts to drop only at a higher $[60/100]$ level, is
evident.  While there is more scatter between galaxies in the NIR/MIR 
ratios at low $[60/100]$ values, the $[2.2/15]$ value does 
(anti)correlate closely with $[60/100]$.
In contrast, the slope of $[2.2/6.7]$ has a break.  
Empirically, examining the lower panels, it is 
clear that the drop in $[6.7/15]$ is caused by the stronger increase 
of 15 $\umu$m emission relative to that at 6.7 $\umu$m.  Also, the 
$[6.7/15]$ and $[2.2/6.7]$ slopes have almost exactly inverted shapes, 
which results in the linear $[2.2/15]$ vs.\ $[60/100]$ relation.

What then do the slopes of colour indices tell us?  
First of all, the nearly constant $[6.7/15]$ 
at $[60/100] < 0.4$ heating regime might imply that both MIR bands are 
dominated here by emission from a common source, namely the UIBs, 
as suggested by Roussel 
\ea (2001b).  On the other hand, since the 11.3 and $12.7 \umu$m UIB bands 
contributing to the LW3 filter are thought to be weak, it might also
be a common {\em site} of the emission, rather than
common physical origin, which is more important.  Nevertheless,
at higher heating levels the global levels of 6.7 and 15 $\umu$m emission
clearly behave differently.  A simple and common interpretation of the
drop in the $[6.7/15]$ ratio has been that after a threshold, 
the heated continuum from very small grains enters the 15 $\umu$m band.
But if this were the {\em only} 
effect one should expect also a break in the 
$[2.2/15]$ slope, which is not seen.  And since the 
strength of PAH emission would be expected to follow the interstellar 
radiation field (ISRF), the break in $[2.2/6.7]$ could be taken to 
indicate the depletion of UIBs in galaxies with hottest $[60/100]$
(see Cesarsky \ea 1996 for the effect in localized intense radiation 
environments). 
The ISO-IRAS diagram must then be explained by both the increasing 
15 $\umu$m emission and decreasing (relative to ISRF) 
6.7 $\umu$m emission.  

Are these effects driven mainly by differing 
amounts of quiescent and active media in the galaxy as a whole 
(as in the two-component model, Helou 1986, 
Dale \ea 1999), by different proportions of star-formation 
happening in the disk vs.\ the central parts 
(\eg Roussel \ea 2001b, Vigroux \ea 1999), or something else?  
To study this in detail is out of scope of the present paper, and in any
case cannot be done only with integrated, global values.
However, it is interesting to note the trends with the morphologies.
The galaxies with constant $[6.7/15]\sim1$ 
are mainly normal disk galaxies, and, in fact, their
spread in $[2.2/15]$ and $[2.2/6.7]$ is correlated with their Hubble type 
as already seen in Section~\ref{class}.  Those with higher $[60/100]$ 
span all morphological types, but stand out by having been classified
as peculiar one way or another (more actice nuclear regions?). 
Thus, it appears, that at lower heating 
levels ($[60/100] \sol 0.4$) the NIR/MIR (and FIR) colours of galaxies 
are driven by morphology, \ie by the spatial distribution of ISM. 
At higher $[60/100]$ the trends on the other hand follow closely the
increasing radiation field, the warming dust continuum, and (possibly)
the destruction of PAH carriers. This is in agreement with the
MIR/FIR studies of Sauvage \& Thuan (1994), who find the FIR colours along
the Hubble sequence to be driven by both star-formation efficiency and
spatial distribution of dust.

\subsubsection{Nuclear star-formation}

As seen in Fig.\ref{fircols}, there are several early type galaxies
in the high $[60/100]$ region of the panels.  These must have
strong nuclear star formation since the MIR and
FIR colours of the galaxies
are totally dominated by a starburst.  However, it is interesting that
the inclusion of near-IR photometry 
distinguishes several galaxies with high NIR/MIR ratios which otherwise 
are tightly placed within the main group of points in the {\em IRAS} 
and {\em ISO-IRAS}-plots.  These are also all early types, but apparently
not true starbursts.  They simultaneously have relatively high heating 
levels (especially one at $[60/100]\approx0.8$, NGC 1266) and high 
NIR/MIR ratios, typical of more normal lenticulars.  This may suggest,
for example, that centrally concentrated dust is heated by the 
high ISRF environment found in the centres of ellipticals and lenticulars 
(Sauvage \& Thuan 1994). However, we also note that all six galaxies
at $[2.2/6.7] > 2$ (after excluding one elliptical) have signs of nuclear 
activity: 5 are LINERs and one has a Sy2 nucleus.  It thus 
seems that the use of the NIR/MIR ratio picks out galaxies with weak 
active nuclei from the MIR/FIR sequence.
Galaxies which are fully dominated by an AGN are expected to have 
a much higher $[60/100]$.  For example, the most extreme source at 
lower right in Fig.~\ref{fircols} is an AGN (NGC 4418; Spoon \ea 2001, 
Roche et al. 1986).

We also note that the four Dale galaxies with the lowest $[6.7/15]$
ratios are all barred early type spirals (SB0 to SBa). They
have on average $[6.7/15] \approx 0.4$, while the overall average
is $\approx 0.6$.  Two out of five of
the strongly barred early type galaxies have quite normal
$[6.7/15]$.  Though the statistics are weak, this fits very well with the
findings of Roussel \ea (2001c), from a different galaxy sample, that in
some early-type barred galaxies there is excess $15 \umu$m flux due to
recent star-formation triggered by bar-driven gas inflows.
Barred galaxies in general however do not seem to have greater
star-forming activity than the non-barred cases.

\subsubsection{Star formation in ELAIS galaxies}

The ELAIS galaxies have on average $[6.7/15] \approx 0.67 \pm 0.27$, which
indicates that the majority of them seem to be star-forming.  While
some of the sources might be at redshifts 
which warrant a significant correction
to acquire the true ratio, most of the objects are expected
to lie at small redshifts, $z<0.3$.  This is strongly suggested by 
Figs.~\ref{6715-vs-k15-ii} and~\ref{6715-vs-k67} (see discussion in 
Section~\ref{colours}).  The same redshift range can also be 
derived from typical NIR and MIR fluxes. 
The median $K$-magnitude of the galaxies
with detections in both mid-IR filters is 14.1 mag which gives an 
expected median redshift of $z \approx 0.15$ (Songaila \ea 1994).
Flores \ea (1999) obtained spectroscopy for a deeper sample of 15 $\umu$m 
ISOCAM galaxies and found a median redshift of $z\sim 0.7$.  Our galaxy
sample is an order of magnitude brighter thus indicating typical 
redshifts of $z\sim 0.2$ or less. 
If only identified LW3 ELAIS galaxies are considered, the
median redshift could maximally be at $z\sim0.3$.
As seen \eg from Fig.~\ref{6715-vs-k15-ii}, K-corrections of Sc's and
starbursts decrease the $[6.7/15]$ ratio by a
factor of $\sim 2$ out to a redshift of $z \approx 0.4$.  
The effect is much smaller
for earlier types.  The redshift-corrected $[6.7/15]$ ratios are thus
likely to stay below the quiescent $[6.7/15]\sim 1$ value.
The lowest detected $[6.7/15]$ are at $\sim 0.3$, which
would indicate significant dust heating: $[60/100] \sim 0.6 - 0.9$,
allowing for redshift effects in the $6.7 \umu$m band.

A rough estimate of star formation rates (SFR) expected can be made 
utilizing
relations in Roussel \ea (2001b). They found a good correlation between
mid-IR emission and $H\alpha$, and thus SFR. 
The correlation holds only in disks of 
spirals, however, or globally only in galaxies where the integrated flux 
is dominated 
by the disk.  From our sample, we selected quiescent and likely 
disk-dominated sources, \ie those with $[6.7/15]$ close to unity and falling 
which fell to the 'Scd' area in our classification.
There are eight such sources, six of which with $K\sim14$ 
mag.  Assigning these a redshift of $z=0.17$
and using blindly the SFR relations from Roussel \ea (2001b), with
assumptions and filterwidths therein, the average SFRs translate to 
$\sim 15 - 30 M_{\odot} / yr$.  Some of the fainter ELAIS galaxies would 
get SFRs
several times this value; however, the application of the relation is 
highly uncertain without more information of the sources. 
The two remaining objects of the selected eight are those labeled 'C' 
and 'D', the bright galaxy pair in Fig.~\ref{nir-early}.
These lie at about $z=0.03$, and would come out with 
SFR($M_{\odot} / yr$) $\approx$ 7 and 3, respectively.

Finally, we note that many of the ELAIS sources (as the pair just mentioned) 
appear to be part of a double or multiple system, 
some with disturbed morphology.  Tidally triggered
star-formation clearly plays an important role in mid-IR studies of 
galaxies.
We have verified this trend with deeper near-IR follow-up observations
of the faintest (and blank field) {\em ISO}-detections using the
IRTF; the results will be discussed elsewhere.

\subsection{Quasars, AGN, and EROs}
\label{qsos}

\begin{figure}
\centerline{\psfig{figure=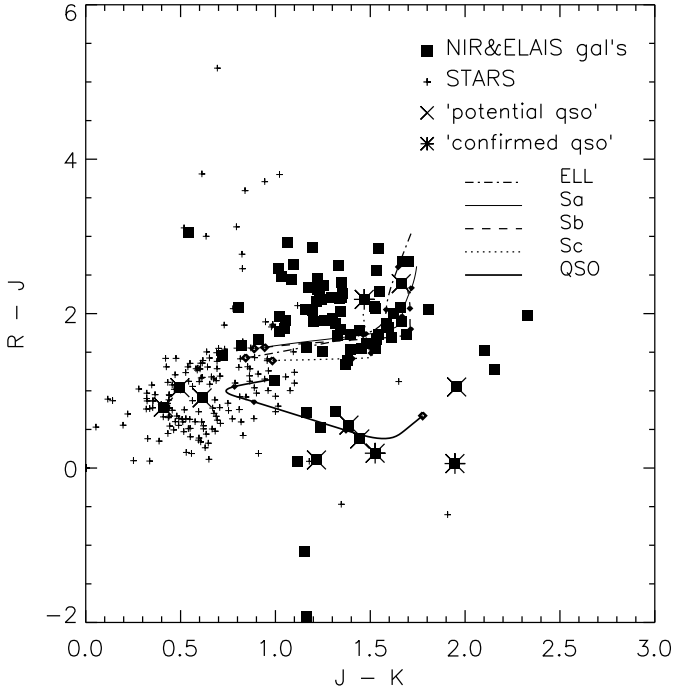,width=10cm}}
\caption{$R-J$ plotted against $J-K$ colour. The solid symbols are the
ISOCAM galaxies (one or both filter detections).
The ISOCAM detections classified
as stars are overplotted as small crosses.
Most stars occupy the lower left region, while dwarf stars rise to
red $R-J$ colours.
Those galaxies, which morphologically seem point-like, or are too faint to
to be classified morphologically from our near-IR data, are overplotted as
large crosses. Three of these turn out to be catalogued quasars,
and are marked with an asterisk (shown also in Figs.~\ref{j-k_plot}
and~\ref{j-k_plot2}).  Models are overplotted as before.
}
\label{j-k_optplot}
\end{figure}

As shown \eg in Laurent \ea (2000), the signature of AGN is a strong
rising continuum starting already at 3 $\umu$m.
Returning to Figs.~\ref{j-k_plot}, \ref{j-k_plot2}, and~\ref{15k-vs-7k},
we further investigated the sources with low NIR/MIR ratios,
in order to check the capability of NIR/MIR for AGN/QSO detection.

Making use of the APS colours, Fig.~\ref{j-k_optplot} shows the $R-J$ colour
(R being the POSS `E' magnitude) against $J-K$.
The solid symbols are the near- and mid-IR ELAIS galaxies, while we have 
also
marked the stars with small symbols.
This plot is equivalent to those used in optical/near-IR searches for
(obscured) QSOs
(the 'KX-method'; see Warren, Hewett \& Foltz 2000; Francis, Whiting \&
Webster 2000; Barkhouse \& Hall 2001).
In general, QSOs tend to have red $J-K$ colours in contrast to a blue $R-J$
(or $B$ or $V$ instead of $R$).

We first selected from our galaxy sample those objects
which had a stellar or ambiguous morphology from APS (20
objects in total; see also Figs.~\ref{relstar_matches1}
and~\ref{relstar_matches2}).  We checked each of these individually from our
near-IR data and many turned out to be galaxies.  Several were,
however, either point-like or too faint to classify: these
`potential QSOs' are overplotted with large diagonal
crosses in Fig~\ref{j-k_optplot}, and also in Figs.~\ref{j-k_plot}
and~\ref{j-k_plot2}. It is quite interesting that the majority of these
do, indeed, fall by the NIR/MIR colours predicted for QSOs/AGN (note that
the selection was done only by optical/near-IR colours and morphology).
All the 5 potentials with $6.7 \umu$m flux have a very steep
$K$ to $6.7 \umu$m gradient.  Since the selection here required a POSS
detection, the faintest NIR objects are excluded -- for example, the
three lowest empty squares in Fig.~\ref{j-k_plot} at $J-K >1.3$ are also
point-like to our resolution, and have near- and mid-IR characteristics
similar to the `potential QSOs'.

Part of our N2 region is covered by Crampton et al. (1988) quasar
catalogue. Three of the `potential QSOs' are found in the catalogue
(and are marked with a large asterisk in the figures mentioned), which
demonstrates the usefulness of the KX-method.  Note that this does not
rule out that other potentials, specifically among those 7 with $J-K > 1.4$,
could not be QSOs due to only a
partial overlap with the quasar catalogue.   The two points at $R-J>2$
are optically faint, near the APS limits, and are brought down to the `QSO
area' when using NIR/MIR.  Three of the potentials have a very
blue, star-like, $J-K$ -- on the other hand they have clear mid-IR excess
as evidenced by LW3.

Oddly, only two of these `potential QSOs' have both ISOCAM
filter data available -- in Figs.~\ref{6715-vs-k15-ii} and~\ref{6715-vs-k67}
these two are the points with the smallest NIR/MIR ratios.
As seen in Fig~\ref{15k-vs-7k}, they are indeed the two ELAIS objects
which fall on the `AGN-area' of our classification system 
(Section~\ref{class}).
The $[6.7/15]$ ratio alone does not separate them from late-type galaxies.
Those of the QSO potentials which have a LW3 detection only, are also likely
candidates for Seyfert 2's, which seem to have a suppressed $6-12 \umu$m
continuum (Spinoglio \ea 1995).  Overall, QSOs have quite a large range of
spectral shapes in the mid-IR (Haas \ea 2000)
making detailed predictions difficult.

The reddest source in Fig.~\ref{j-k_plot}
has $J-K \approx 2.5$, which would qualify it as a 
extremely red object (ERO) candidate
(eg.\ Cimatti \ea 1999; Scodeggio \& Silva 2000; Pozzetti \& Mannucci 2000)
-- it does not have any optical counterpart to the POSS limits, making it
at least $R-K > 5$.  It has only a LW3 detection, but this is due to lacking
coverage with LW2.  Thus far EROs have been selected and studied in the 
optical
and near-IR, and there is an interesting degeneracy in explaining their
nature:  their colours could signify either an old elliptical
at $z>1$, or a young, dusty star-forming galaxy.  According to the
GRASIL models, an elliptical would have $J-K \sim 2.5$
at $z > 2$.  Ellipticals, especially distant ones,
would not have been seen by the ELAIS survey, and in any case
the low $[2.2/15] \approx 0.1$ colour shows the presence of
significant dust emission.
Thus, the mid-IR can break the degeneracy of ERO observations.  A detailed
search for red objects using deeper optical imaging and our near-IR data,
accompanied with the mid-IR {\em ISO}-data, is thus of high importance.

\section{Conclusions}

1.
We have presented photometry of a subsample of the ISOCAM ELAIS survey from 
the
N1 and N2 fields.  Our near-IR survey reaches down to $J \approx 19$ and
$K \approx 17.5$.  All of the $6.7 \umu$m (LW2)
REL=2 sources are identified to these limits,
as well as 84 per cent of $15 \umu$m (LW3) REL=2 sources.
The detection efficiencies for REL=3 sources are 88 and 35 per cent
at LW2 and LW3 bands, respectively.

2. The near- and mid-IR stars were used, along with stellar models, 
to perform an accurate new
calibration of the ELAIS ISOCAM data at both 6.7 and 15 $\umu$m.

3.
Stars were separated from galaxies using near- to mid-IR colours.
At 6.7 $\umu$m, 80 per cent of the identified ELAIS objects are stars.  
In contrast,
at 15 $\umu$m, 80 per cent of the near-IR identified ELAIS sources are 
galaxies.

4. 
Only one third of LW3 galaxies are also detected in LW2, while two thirds of 
LW2 galaxies are seen in LW3. 
The mid-IR survey as a whole mainly detects late type spiral galaxies
and starbursts. The
faintest population of these is missed by the LW2 filter. The few objects 
missed by the longer mid-IR filter are most
probably early type galaxies.  Simple arguments indicate that
typical redshifts of the sample seen with both mid-IR bands
are $z \leqslant 0.2$.

5.
We have presented several colour-colour plots useful in studying the 
relative
emission strengths of stellar, PAH, and warm dust components in galaxies
and 
we discuss galaxy classification and star formation properties using the
diagrams.  In a $[15/2.2]$ vs.\ $[6.7/2.2]$ plot
the Hubble type of a galaxy can be roughly estimated from its position
along the diagonal ($[6.7/15] = 1$), which is a measure of the
proportion of ISM in the galaxy.   Of the near-IR-identified 
galaxies detected with both mid-IR filters, 75 per cent fall in the Scd-group.
However, some of these might be earlier morphological types with significant
nuclear star formation.

6.
In the same $[15/2.2]$ vs.\ $[6.7/2.2]$ plot the quiescent
galaxies fall on the diagonal (where $[15/6.7] \approx 1$)
with increasing star formation activity raising the galaxies above
the one-to-one curve.  The ELAIS galaxies are found to have
significant star-formation, as indicated by the $[6.7/15]$ tracer
($f_{\nu}(6.7\umu{\rm m}) / f_{\nu}(15\umu{\rm m}) = 0.67 \pm 0.27$)
as well as by estimates from published relations between mid-IR luminosity
and SFR.
Redshift information and resolved imaging
is however needed to better quantify SFRs and to decide whether 
the ELAIS galaxies are
powered by strong nuclear starbursts or otherwise high star formation 
activity in the disk.

7.
In quiescent galaxies, as indicated by their $[60/100]$ IRAS colour, 
$[6.7/15]$  remains very constant.  These are also the galaxies where
the classification of galaxies using NIR/MIR ratios works the best.  
The MIR ratio starts to drop at hotter $[60/100]$.
Using NIR/MIR colours we find support 
for the view that both the increase of $15\umu$m emission and an
apparent depletion of emission at $6.7\umu$m 
are responsible for the effect.  At these higher $[60/100]$ levels,
both $[6.7/15]$ and $[2.2/15]$ ratios
(anti)correlate well with the $[60/100]$ activity level indicator, 
thus making them useful tracers of star-formation. 

8.
The ELAIS survey covered here detects several active galactic nuclei.
By selecting objects using a `KX-method'
(considering optical to near-IR properties only) we pick out sources
from our
catalogue, whose mid-IR fluxes are consistent with the objects being
AGN/QSOs.

\begin{table*}
\begin{center}
\small % \tiny
\caption{The near-IR sample of ELAIS galaxies detected with both ISOCAM 
filters.  The galaxies are ordered with decreasing $15 \umu$m flux. 
Columns 7 and 9 labeled `R' refer to the REL parameter.  
The coordinates (J2000) are from the NIR data.
The bright galaxies A to E from Fig.~\ref{nir-early} are indicated; `fir'
and `vla' indicate that the objects has been detected in the 90 $\umu$m
ELAIS survey (Efstathiou \ea 2000) and a VLA followup survey 
(Ciliegi \ea 1999); `q1' and `q2' indicate a confirmed and potential quasar,
respectively, as discussed in Section~\ref{qsos}.
} 
\label{table1}
\begin{tabular}{lccccccccccl}
\hline
  &  RA & DEC & $J$ mag & $J$-err& $K$ mag & $K$-err & $S_{15}$ (mJy) & R & $S_{6.7}$ 
(mJy) & R & {Notes} \\
\hline
  1  &  16 37 34.4 & +40 52 08 & 13.22  &  0.02 &  12.20  &  0.02  &  34.3  &   2  &  37.3  &   2 & 
   `C', vla  \\
  2  &  16 35 07.9 & +40 59 29 & 12.65  &  0.01 &  11.29  &  0.01  &  23.0  &   2  &  22.2  &   2 & 
    `B', fir, vla \\
  3  &  16 34 01.8 & +41 20 52 & 13.06  &  0.01 &  12.03  &  0.02  &  18.5  &   2  &  11.8  &   2 & 
   `E', fir, vla\\ 
  4  &  16 37 29.3 & +40 52 49 & 12.66  &  0.01 &  11.56  &  0.01  &  12.9  &   2  &  18.9  &   2 & 
   `D', fir, vla  \\
  5  &  16 35 25.2 & +40 55 43 & 14.14  &  0.03 &  12.99  &  0.03  &   8.8  &   2  &   9.0  &   2 & 
   fir, vla \\
  6  &  16 37 05.1 & +41 31 55 & 15.26  &  0.04 &  14.02  &  0.04  &   8.5  &   2  &   3.8  &   2 & 
   fir, vla \\
  7  &  16 33 59.1 & +40 53 04 & 14.80  &  0.04 &  13.57  &  0.04  &   7.6  &   2  &   5.3  &   2 & 
  vla\\
  8  &  16 36 08.1 & +41 05 08 & 15.51  &  0.03 &  13.85  &  0.04  &   7.6  &   2  &   2.7  &   2 & 
   vla \\
  9  &  16 35 06.1 & +41 10 38 & 15.48  &  0.04 &  14.10  &  0.03  &   6.2  &   2  &   4.9  &   2 & 
  fir, vla \\
10  &   16 35 46.9 & +40 39 01 & 15.35  &  0.04 &  14.18  &  0.05  &   5.6  &   2  &   3.0  &   2 & 
   vla \\
11  &   16 36 45.0 & +41 51 32 & 14.23  &  0.04 &  13.15  &  0.05  &   4.9  &   2  &   1.7  &   2 & 
 \\
12  &   16 36 13.6 & +40 42 30 & 13.44  &  0.02 &  12.42  &  0.02  &   4.5  &   2  &   3.1  &   2 & 
   vla \\
13  &   16 36 07.6 & +40 55 48 & 15.83  &  0.05 &  14.30  &  0.05  &   4.1  &   2  &   1.3  &   2 & 
  fir, vla \\
14  &   16 37 08.1 & +41 28 56 & 15.26  &  0.04 &  13.91  &  0.04  &   3.6  &   2  &   2.1  &   3 & 
   vla \\
15  &   16 37 31.4 & +40 51 56 & 15.47  &  0.04 &  14.13  &  0.06  &   3.3  &   2  &   1.7  &   2 & 
  fir, vla \\
16  &   16 34 14.2 & +41 03 19 & 14.79  &  0.03 &  13.98  &  0.04  &   2.9  &   2  &   1.5  &   3 & 
   vla \\
17  &   16 34 12.0 & +40 56 53 & 17.59  &  0.11 &  15.43  &  0.08  &   2.9  &   2  &   1.4  &   2 & 
   vla\\
18  &   16 37 20.5 & +41 11 21 & 12.92  &  0.01 &  11.67  &  0.01  &   2.9  &   2  &   2.7  &   2 & 
   `A' \\
19  &   16 36 09.7 & +41 00 18 & 15.89  &  0.05 &  14.45  &  0.07  &   2.9  &   2  &   1.5  &   2 & 
   vla \\
20  &   16 35 19.2 & +40 55 57 & 16.27  &  0.07 &  14.90  &  0.07  &   2.7  &   3  &   0.9  &   3 & 
   vla \\
21  &   16 38 51.9 & +41 10 53 & 15.80  &  0.08 &  14.27  &  0.07  &   2.7  &   2  &   2.1  &   3 & 
 \\
22  &   16 34 23.9 & +40 54 10 & 15.94  &  0.06 &  14.27  &  0.07  &   2.6  &   2  &   1.6  &   2 & 
   vla \\
23  &   16 37 16.8 & +40 48 26 & 14.99  &  0.03 &  13.97  &  0.04  &   2.6  &   2  &   2.1  &   2 & 
   vla \\
24  &   16 35 34.0 & +40 40 25 & 18.02  &  0.15 &  16.06  &  0.15  &   2.3  &   2  &   2.1  &   2 & 
   q2 \\
25  &   16 34 49.6 & +41 20 50 & 15.62  &  0.04 &  ...    &  ...   &   2.2  &   2  &   1.6  &   2 & 
 \\
26  &   16 35 31.2 & +41 00 28 & 17.87  &  0.17 &  16.34  &  0.17  &   2.2  &   3  &   0.8  &   3 & 
   q1 \\
27  &   16 36 40.0 & +40 55 38 & 16.30  &  0.06 &  14.96  &  0.06  &   1.9  &   2  &   1.3  &   2 & 
 \\
28  &   16 34 43.0 & +41 09 49 & 15.60  &  0.05 &  14.30  &  0.06  &   1.9  &   3  &   1.7  &   2 & 
 \\
29  &   16 36 15.2 & +41 19 12 & 16.52  &  0.08 &  15.07  &  0.07  &   1.7  &   3  &   0.9  &   3 & 
\\
\hline
\end{tabular}
\end{center}
\end{table*}

\section{Acknowledgements}

We wish to thank Kalevi Mattila for very useful discussions and suggestions,
and an anonymous referee for thoughtful and valuable criticism. We thank
A. Boselli for providing his mid-IR fluxes.
This research has made use of the NASA/IPAC Extragalactic Database (NED)
which is operated by the Jet
             Propulsion Laboratory, California Institute of Technology,
under contract with the National Aeronautics and
             Space Administration.
This publication makes use of data products from the Two Micron All Sky
Survey, which is a joint project of the University of Massachusetts and the
Infrared Processing and Analysis Center/California Institute of Technology,
funded by the National Aeronautics and Space Administration and the National
Science Foundation.
This research has made use of the APS Catalog of POSS I,
which is supported by the National
            Aeronautics and Space Administration and and the
            University of Minnesota. The APS databases can be
            accessed at http://aps.umn.edu/.

\appendix

\section{Calibration of ISOCAM fluxes using infrared stars}
\label{appdx}

The ELAIS catalogue v.1.3 uses a one to one conversion of ADUs/gain/s to mJy
fluxes.  The ISOCAM handbook values are 2.32 and 1.96 ADU/gain/s/mJy for
the 6.7 and $15 \umu$m filters, respectively (Blommaert 1998) --
the reason for a factor of $\sim 2$ difference is the lack of
source stabilization correction in ELAIS data
as detailed in Serjeant et al. (2000) (see also Blommaert 1998 -- it is 
noted
therein, that the stabilization correction remains the largest single
uncertainty in ISOCAM flux calibration).
In other words, starting from the handbook value of $\sim 2$ ADU/gain/s/mJy
and correcting for the loss of flux resulting from lack of stabilization,
the conversion becomes $\sim 1$ ADU/gain/s/mJy.

However, in case of the LW3 data Serjeant
et al. (2000) find a discrepancy of a factor of 1.75 after a 
cross-correlation
with 22 bright stars in the ELAIS fields -- in that paper all 15$\umu$m 
fluxes
are thus multiplied by a factor of 2.  The publicly available v.1.4 ELAIS
catalogue uses the factor of 1.75 in $15 \umu$m fluxes.
In Missoulis et al. (1999)
mid-infrared fluxes were derived for the same stars using
B and V-band bolometric magnitudes from Hipparcos and SIMBAD, along with
blackbody approximations.  Correlating with observed fluxes,
sensitivity factors of 0.56 and 0.70 ADU/gain/s/mJy were obtained for
6.7 and $15 \umu$m, respectively.

With good-quality near-IR data, rather than optical data,
we potentially have a better chance of
deriving the calibration factor for ELAIS data
using the stars in our survey area.  We would
greatly reduce the uncertainty of extrapolating
the optical magnitudes into mid-IR, as well as the
required precision in the spectral types of stars.

To compare with observations, we make use of observationally based stellar
spectra used for the extensive ISOCAM and ISOPHOT calibration
programs\footnote{see 
http://www.iso.vilspa.esa.es/users/expl\_lib/ISO/wwwcal/ \\ cam.html/}.
We calculated near- and mid-IR colours of stars with a range of
spectral types from these spectra.  The models
are estimated to be accurate within 5 per cent.
In the mid-IR the fluxes were
colour-corrected (maximally a 7 per cent effect) following the convention
of {\em ISO}-fluxes which are determined using a constant energy spectrum 
(note
that for LW3 the `reference wavelength' is 14.3 $\umu$m).

\begin{figure*}
\centerline{\psfig{figure=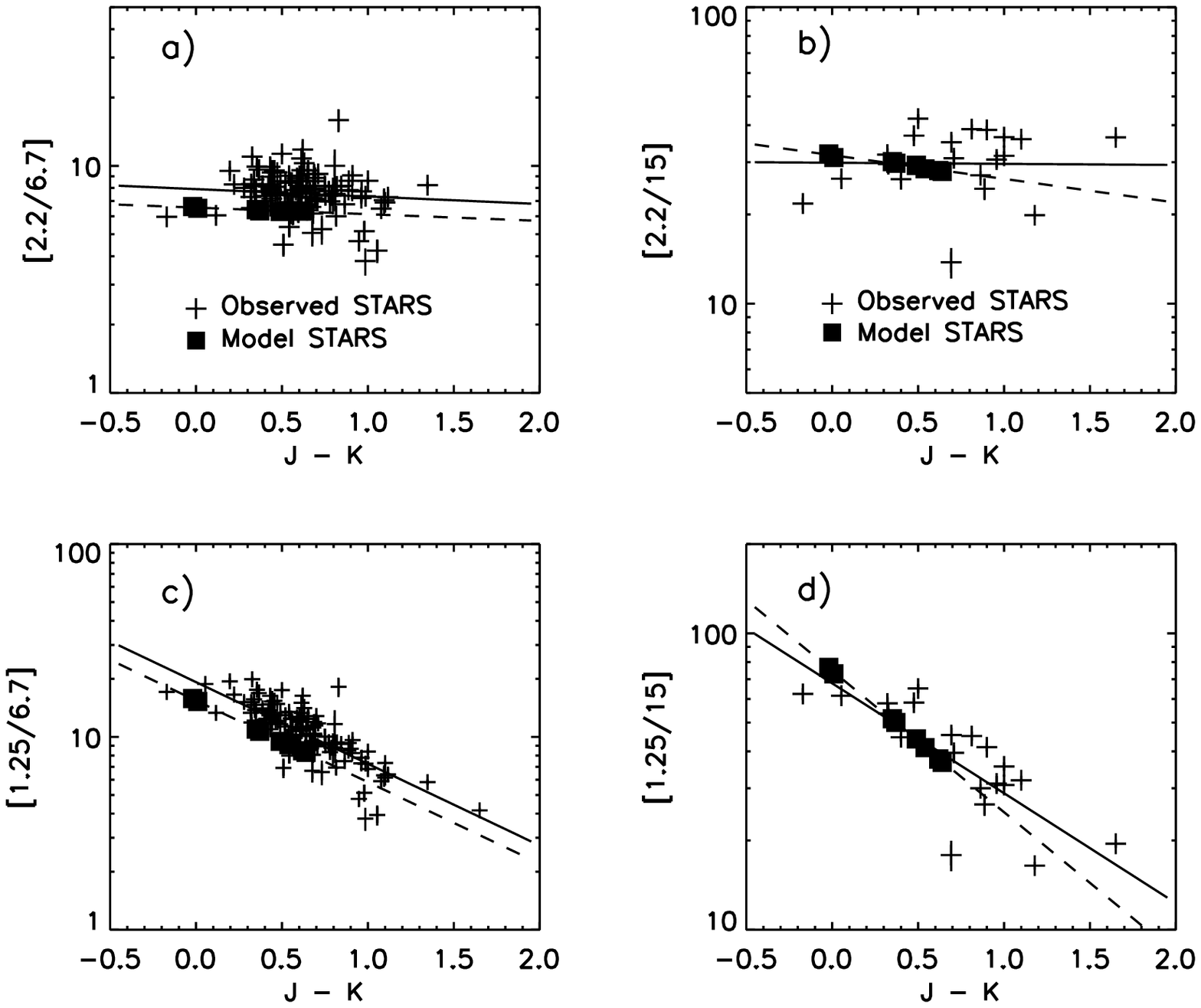,width=13cm}}
\caption{Stars in NIR/MIR vs.\ $J-K$ diagram.
The 6.7 and 15 $\umu$m
fluxes of the ELAIS v.1.3 catalogue stars (crosses)
have been converted assuming 1 ADU/gain/s to 1 mJy.  The $K$-band flux
uses $f_K = 6.20 \cdot 10^5 \times 10^{-0.4\cdot K}$ mJy and the $J$-band
$f_J = 1.52 \cdot 10^6 \times 10^{-0.4 \cdot J}$ mJy.  The model colours
(filled squares) have been calculated from several stellar spectra
templates used in the
ISOCAM calibration program.  The stars range from A0 to K3 in spectral type,
including giants and main-sequence stars.  The reddest of our observed stars
in $J-K$
would be expected to be M-stars. The solid and dashed lines are fits to
the observed data and model points,
respectively.  From these we derive a constant correction factor of
1.22 to the $6.7 \umu$m fluxes (i.e.\ 1.22 ADU/gain/s/mJy) in panel {\em a}.
From panel {\em b}
the conversion of $15 \umu$m flux becomes 1.05 ADU/gain/s/mJy.  Panels {\em 
c}
and {\em d} show the fits using $J$-band magnitudes, instead of $K$, and
the flux calibrations become 1.24 ADU/gain/s/mJy and 1.06 ADU/gain/s/mJy, 
for
the LW2 and LW3 filters, respectively.
}
\label{lw2_stars}
\end{figure*}

From our own sample of stars, defined in Section~\ref{stargal},
we use only those with the REL=2 status.  In addition, we exclude
stars which have $K < 8$ mag, because of probable saturation in our
near-IR images.  Fig.~\ref{lw2_stars}$a$
shows the stars detected at $6.7 \umu$m plotted as $[2.2/6.7]$ vs.\ $J-K$,
with the model stars overplotted as solid symbols.  From the model points
one can notice a slight colour-term, where the later spectral types with
redder $J-K$ have slightly lower $[2.2/6.7]$.
Ignoring the negligible colour-term, from the average difference of
$[2.2/6.7]$ ratios of observations and models, we derive a
correction of 1.22 to the $6.7\umu$m fluxes of the v.1.3 ELAIS catalogue.
Fig.~\ref{lw2_stars}$b$ shows the equivalent plot for the $15 \umu$m stars 
--
there are much less stars here, but the overall calibration of the v.1.3
ELAIS catalogue seems quite accurate.  We derive a  1.05 ADU/gain/s/mJy
calibration for the LW3 data.  Specifically, we do {\em not}
find evidence for the factor of 2 (or 1.75) scaling used in Serjeant et al.\
(2000).  Since we are using the same ELAIS data, from the same reduction
process and the same photometric aperture corrections, the discrepancy has
to come from the adopted method of extrapolating near-IR (our case) or
optical magnitudes to the mid-IR.
The $J$-band data can be used as well: panels $c$ and $d$
show the equivalent colour-colour plots with $J$-flux.  The
calibration factors are confirmed, as we find 1.24 and 1.06
ADU/gain/s/mJy for the LW2 and LW3 filters, respectively.

\begin{figure*}
\centerline{\psfig{figure=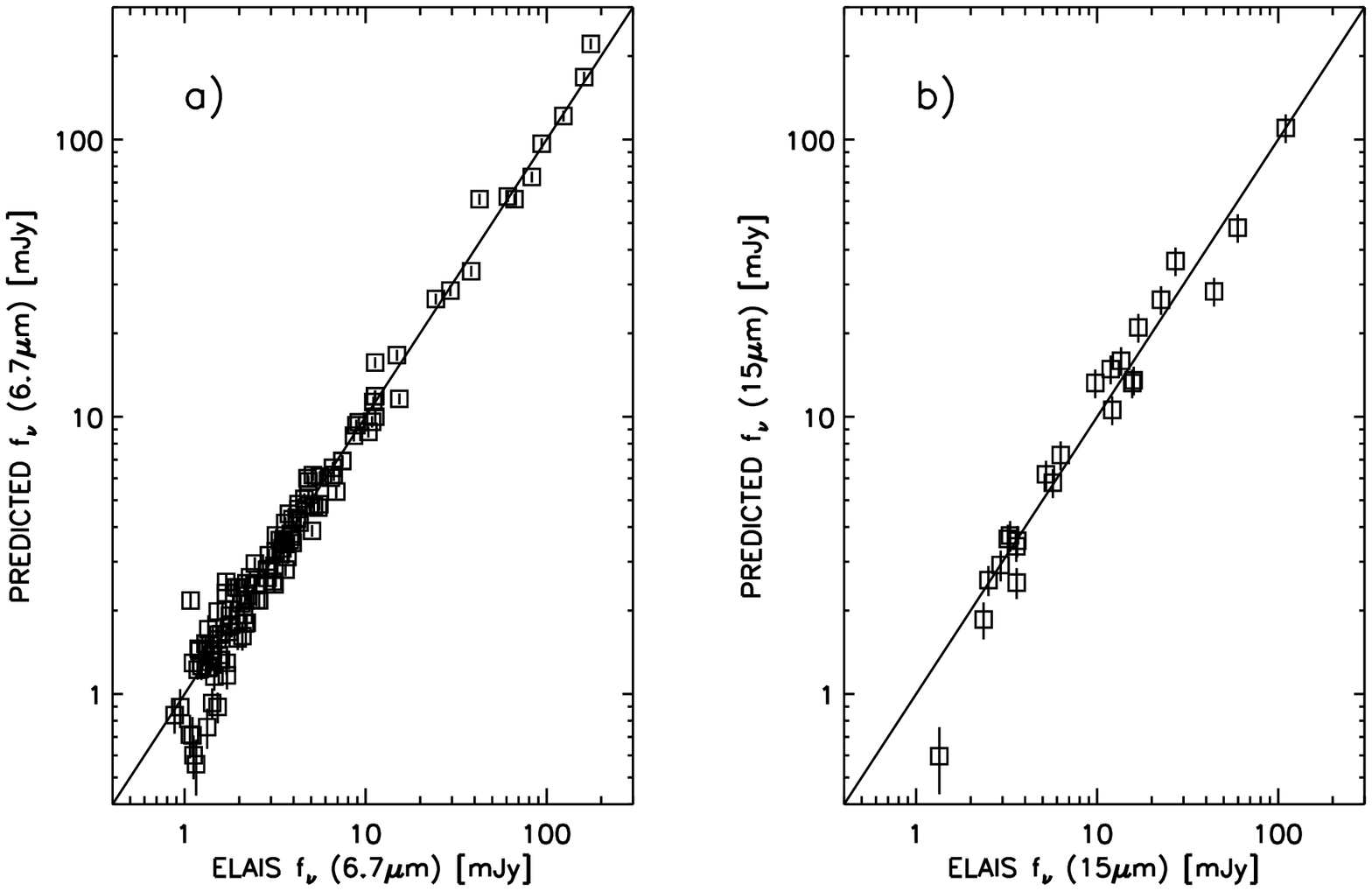,width=11cm}}
\caption{Predicted stellar fluxes vs.\ the observed ELAIS fluxes.
Panel $a$ is for stars at 6.7 $\umu$m and $b$ for the 15 $\umu$m stars.
The predicted flux
derivation uses $K$-band fluxes of stars and the $[2.2/6.7]$ or the 
$[2.2/15]$
ratio found from stellar models.  The observed
fluxes in $a$ and $b$ have been calibrated using the
(small) differences (factors of 1.22 and 1.05, respectively) between the 
model
and observed ratios (see the top panels of Fig.~\ref{lw2_stars}).
}
\label{lw2_stars_corr}
\end{figure*}

To compare with figures in Missoulis et al.\ (1999)
and Serjeant et al. (2000),
Fig.~\ref{lw2_stars_corr} shows the {\em predicted}
6.7 and $15 \umu$m stellar fluxes
(derived from the observed $K$-magnitude of the star
using the corresponding model colour ratio) against the observed
and re-calibrated ELAIS 6.7 and $15 \umu$m fluxes.
The scatter is seen to be very small, and the relation highly linear over 
two
orders of magnitude.  We are thus confident of an
accurate calibration for the ELAIS ISOCAM data.

In summary, in this paper we use the catalogue v.1.3 values for LW2 and
LW3 multiplied by 1.23 and 1.05, respectively, to have the values in mJy
(averages from $K$ and $J$ determination taken).
The correction to conversion for LW3 is, in fact,
smaller than the uncertainties related to the observed spread
in mir-IR fluxes and the models,
but we use it for consistency.  Note that the v.1.4 ELAIS
catalogue has the LW3 fluxes multiplied by 1.75, which needs to be taken 
into
account if compared with results and plots in this paper.

\bsp

\label{lastpage}

\end{document}